\documentclass[12pt]{article}

\usepackage{amssymb,esvect,amsmath,graphicx,latexsym,amsthm,slashed,eso-pic,hyperref}
\usepackage[final]{pdfpages}

\usepackage{epsfig,amsmath,amssymb,verbatim,mathrsfs,hyperref}
\usepackage{xspace}
\usepackage{xcolor}
\usepackage{slashed}
\usepackage{dcolumn}
\usepackage{multirow}
\graphicspath{{./}{Figs/}}

\def\beq{\begin{eqnarray}}
\def\eeq{\end{eqnarray}}
\def\bea{\begin{eqnarray}}
\def\eea{\end{eqnarray}}

\def\mev{\, {\rm MeV}}

\newcommand{\gsim}{\lower.7ex\hbox{$\;\stackrel{\textstyle>}{\sim}\;$}}
\newcommand{\lsim}{\lower.7ex\hbox{$\;\stackrel{\textstyle<}{\sim}\;$}}

\newcommand{\nnmb}{\nonumber}

\newcommand{\lrf}[2]{\left(\frac{#1}{#2}\right)}

\newcommand{\grad}{\vec{\nabla}}

\newcommand{\lag}{\mathscr{L}}

\newcommand{\kev}{\mathrm{keV}}

\newcommand{\beg}{{^8\textrm{Be}}}
\newcommand{\bes}{{^8\textrm{Be}}^*}
\newcommand{\bet}{{^8\textrm{Be}}^{*\prime}}

\newcommand{\ehat}{\hat{e}}
\newcommand{\yhat}[3]{\hat{\mathcal{Y}}_{#1,#3 1}^{#2}}

\definecolor{green1}{RGB}{0,105,0}
\definecolor{cyan1}{RGB}{0,250,250}
\definecolor{magenta1}{RGB}{250,0,250}

\begin{document}
\begin{titlepage}
\noindent
\begin{center}
  \begin{Large}
    \begin{bf}
Light Axial Vectors, Nuclear Transitions, \\ and the $^8$Be Anomaly
     \end{bf}
  \end{Large}
\end{center}
\vspace{0.2cm}
\begin{center}
\begin{large}
Jonathan Kozaczuk$^{(a,b)}$, David E. Morrissey$^{(b)}$, and S.~R.~Stroberg$^{(b,c)}$
\end{large}
\vspace{1cm}\\
\begin{it}
(a) Amherst Center for Fundamental Interactions, Department of Physics,\\ 
University of Massachusetts, Amherst, MA 01003, USA
\vspace{0.2cm}\\
(b) TRIUMF, 4004 Wesbrook Mall, Vancouver, BC V6T 2A3, Canada
\vspace{0.2cm}\\
(c) Reed College, 3203 SE Woodstock Blvd, Portland, OR 97202, USA 
\vspace{0.5cm}\\
email: \emph{\texttt{kozaczuk@umass.edu}},
\emph{\texttt{dmorri@triumf.ca}}, 
\emph{\texttt{sstroberg@triumf.ca}}
\vspace{0.2cm}
\end{it}
\end{center}
\center{\today}

\begin{abstract}

New hidden particles could potentially be emitted and discovered 
in rare nuclear transitions.  In this work we investigate the production 
of hidden vector bosons with primarily axial couplings to light quarks in 
nuclear transitions, and we apply our results 
to the recent anomaly seen in $^8$Be decays.  The relevant 
matrix elements for $\bes(1^+) \to \beg(0^+)$ transitions 
are calculated using \emph{ab initio} methods with 
inter-nucleon forces derived from chiral effective field theory 
and the in-medium similarity renormalization group.  
We find that the emission of a light axial vector 
with mass $m_X \simeq 17\,\mev$ can account for the anomaly seen 
in the $1^+\to 0^+$ isoscalar transition 
together with the absence of a significant anomaly in the corresponding 
isovector transition.  We also show that such an axial vector can be
derived from an anomaly-free ultraviolet-complete theory that is consistent
with current experimental data.

\end{abstract}

\end{titlepage}

\setcounter{page}{2}


\section{Introduction\label{sec:intro}}

  The search for new forces has been a longstanding pursuit of subatomic 
physics research~\cite{Leike:1998wr,Langacker:2008yv}.  
New force carriers that couple significantly to the Standard Model~(SM)
have been searched for directly at high energy colliders 
such as the LHC~\cite{Carena:2004xs,Rizzo:2006nw,Aaboud:2016cth,Khachatryan:2016zqb} and tested indirectly through high-precision 
measurements~\cite{Erler:2009jh}, and they must have masses well 
above the electroweak scale to be consistent with these data.  
Exotic force carriers with masses below the electroweak scale are
also allowed by current experiments if they are \emph{hidden}, coupling
very weakly to SM matter~\cite{Borodatchenkova:2005ct,Fayet:2007ua,Pospelov:2008zw,Bjorken:2009mm}.  
The most sensitive searches for light hidden states
are typically lower-energy collider experiments with a very high intensity
of collisions~\cite{Bjorken:2009mm,Batell:2009yf,Essig:2009nc,Reece:2009un,Essig:2013lka,Alexander:2016aln}.  
Experiments of very high precision are also competitive in terms 
of current limits and future discovery 
prospects~\cite{Pospelov:2008zw,Essig:2013lka}.

  Light vector boson force carriers and other light hidden particles 
with masses up to a few tens of MeV can also be searched for in rare nuclear 
decays~\cite{Donnelly:1978ty,Treiman:1978ge}.  
Various types of hidden particles can be emitted in such transitions 
depending on the spin and parity of the initial and final nuclear states.  
Indeed, significant limits on axions have been derived from precision 
measurements of $^8$Be, $^{14}$N, and $^{16}$O decays~\cite{Savage:1986ty,Hallin:1986gh,Savage:1988rg}.
More recently, the emission of hidden vector bosons by nuclei has received 
particular attention due to an apparent anomaly seen in measurements 
of $^8$Be transitions~\cite{Krasznahorkay:2015iga}.  

An experiment at the MTA-Atomki facility reports a significant ($6.8\sigma$)
bump in the distribution of opening angles between energetic electron-positron 
pairs emitted in isoscalar $\bes(1^+) \to \beg(0^+)+e^+e^-$ 
transitions~\cite{Krasznahorkay:2015iga}.  No such bump is expected
from known nuclear physics, which predicts that this transition arises
primarily from internal pair conversion with a smoothly falling
distribution of $e^+e^-$ opening angles.  Furthermore,
no significant excess is seen in the related isovector 
$\bet(1^+)\to \beg(0^+)+e^+e^-$ transition~\cite{Krasznahorkay:2015iga}.  
For future reference, we list the relevant $\beg$ states in Table~\ref{tab:be8},
together with their masses, excitation energies, relevant decay widths, 
and angular momentum~($J$), parity~($P$), and approximate isospin~($T$) 
quantum numbers~\cite{Tilley:2004zz}.

This apparent anomaly in $\beg$ transitions can be explained by 
an additional decay channel to a light vector boson $X$, 
$\bes(1^+) \to \beg(0^+)+X$,
followed by $X\to e^+e^-$~\cite{Krasznahorkay:2015iga,Feng:2016jff}. 
To match the kinematic feature seen in $e^+e^-$ opening angles, the new
vector should have a mass $m_X\simeq 17\,\mev$~\cite{Krasznahorkay:2015iga}.
This proposal was studied in detail in Refs.~\cite{Feng:2016jff,Feng:2016ysn}
for a vector boson with purely \emph{vector} (as opposed to axial) couplings
to quarks.  These works showed that such an explanation can be
consistent with existing experimental constraints provided the new vector 
is approximately \emph{protophobic}~\cite{Feng:2016jff}, 
coupling much more weakly to the proton than to the neutron.  
Further related investigations and interpretations of the excess 
have appeared as well~\cite{Gu:2016ege,Chen:2016dhm,Liang:2016ffe,Jia:2016uxs,Kitahara:2016zyb,Ellwanger:2016wfe,Chen:2016tdz,Kahn:2016vjr,Seto:2016pks,Neves:2016ugb}.    
  
  In this work we investigate whether a new vector boson with primarily
\emph{axial} couplings to quarks can account
for the $^8$Be anomaly.  This possibility was suggested in 
Refs.~\cite{Feng:2016jff,Feng:2016ysn}, but it was not pursued systematically
due to the difficulty of computing the corresponding nuclear matrix elements.
We confront this challenge head on, and apply state-of-the-art 
\emph{ab initio} nuclear theory methods to derive a controlled estimate 
of the relevant nuclear physics quantities.  We then apply our results 
to the $^8$Be anomaly to determine whether a hidden axial vector 
can provide a viable explanation.

The outline of this paper is as follows.  After this introduction, 
we adapt the formalism of electromagnetic and weak nuclear decays
to general nuclear decays with the emission of a light hidden particle
in Section~\ref{sec:ndecay}, and we apply it to the $\beg$ system with
a light vector with axial couplings to quarks.  In Section~\ref{sec:nuc}
we present our nuclear physics calculation of the transition matrix elements
relevant to the $^8$Be anomaly.  These results are then applied to
study an axial vector interpretation of the anomaly in Section~\ref{sec:be8}.
A comparison of this interpretation with other limits on light axial
vectors are studied in Section~\ref{sec:constraints}.  
We comment on UV completions with light axial vectors consistent with 
the $^8$Be anomaly in Section~\ref{sec:UV}.
Finally, Section~\ref{sec:conc} is reserved for our conclusions.

\begin{table}[ttt]
\begin{center}
\begin{tabular}{l|c|c|c|c|c}
{State}&$m~(\mev)$&$\Delta E~(\mev)$&$\Gamma~(\kev)$&$\Gamma_{\gamma}~(\mathrm{eV})$&$J^P_T$\\
\hline
$\beg$&7454.85&0&{--}&{--}&$0^+_0$\\
$\bes$&7473.00&18.15&138&1.9&$1^+_0$\\
$\bet$&7472.49&17.64&10.7&15&$1^+_1$\\
\end{tabular}
\end{center}
\caption{$\beg$ ground and excited states relevant to the Atomki 
anomaly~\cite{Krasznahorkay:2015iga} together with their mass, 
excitation energy, total decay width, decay width to $\beg+\gamma$,
spin~($J$), parity~($P$), and approximate isospin~($T$) 
assigments~\cite{Tilley:2004zz,Feng:2016jff}.
\label{tab:be8}}
\end{table}

\section{Nuclear Decay Rates to a Massive Vector\label{sec:ndecay}}

  In this section we adapt the formalism of electromagnetic and weak nuclear
decays to general nuclear transitions in which a light (but massive) vector
boson is emitted, and we derive a general expression for the corresponding
decay rate in terms of the underlying nucleon current coupling.  
Next, we specialize to a light vector with axial couplings to quarks
and obtain the relevant nucleon-level currents and a simplified
expression for the transition matrix elements. 
These results are then applied to $\bes(1^+)\to \beg(0^+)$ transitions.

\subsection{General Formalism for Nuclear Decays}

  Consider a massive vector boson $X$ that couples to hadrons in the
SM through the current
\beq
\mathcal{H}_{int} ~\supset~ \mathcal{J}_{\mu}X^{\mu} \ .
\label{eq:vcoup}
\eeq
This interaction can potentially lead to nuclear decays of the form 
$\vert i\rangle \to \vert f\rangle + X$, provided the vector 
is light enough. At leading order in the interaction of Eq.~\eqref{eq:vcoup},
the corresponding (Schr\"odinger picture) transition matrix element is
\beq \label{eq:M}
\mathcal{M} = \int\!d^3x\,\langle f\vert\mathcal{J}_{\mu}\,\epsilon^{\mu\,*}_{a}
e^{-i\vec{k}\cdot\vec{x}}\,\vert i\rangle \ ,
\eeq
where $\epsilon^{\mu}_{a}$ is the polarization vector of
the outgoing vector boson with 3-momentum $\vec{k}$ and polarization
state $a$.  

To evaluate the nuclear matrix element, it is conventional to expand it 
in terms of spherical tensor operators~\cite{Blatt,Walecka:1995mi}.  
If the initial state is unpolarized, the quantization axis for 
angular momentum can be chosen parallel to $\vec{k} \to k\,\hat{z}$.  
In this case, the three polarization vectors can be taken to be
\beq
\epsilon^{\mu}_0 = \frac{1}{m_X}(k,0,0,E_k) \ ,~~~~~
\epsilon^{\mu}_{\pm 1} = \mp\frac{1}{\sqrt{2}}(0,1,\pm i,0) \ .
\eeq
Defining the spherical basis $\ehat_0 = \hat{z}$ and 
$\ehat_{\pm 1} = \mp(\hat{x}\pm i\hat{y})/\sqrt{2}$, we have
\beq
\vec{\epsilon}_a^{\;*} = \sum_{\lambda}(\vec{\epsilon}^{\;*}_a\cdot\hat{e}_{\lambda})
\hat{e}_{\lambda}^* ~\equiv~ \sum_{\lambda}(\epsilon^*_a)_{\lambda}\,
\hat{e}_{\lambda}^*
\eeq
with
\beq
(\epsilon^*_a)_{0} ~=~ \frac{E_k}{m_X}\delta_{a0} \ ,~~~~
(\epsilon^*_a)_{\pm 1} ~=~ \delta_{a\pm 1} \ ,~~~~ 
\epsilon^{0\,*}_a ~=~ \frac{k}{m_X}\delta_{a0} \ . 
\label{eq:eps}
\eeq
The operators $\ehat_{\lambda}^*e^{-ikz}$ can be expanded in a spherical vector
basis to give~\cite{Walecka:1995mi}
\beq
\mathcal{M} &=& -\langle f\vert\left(
\sum_{J\geq 1}\,(-i)^J\sqrt{2\pi(2J+1)}\sum_{\lambda=\pm 1}(\epsilon^*_a)_{\lambda}
\left[\lambda\mathcal{T}^{mag}_{J,-\lambda}(k) 
+ \mathcal{T}^{el}_{J,-\lambda}(k)\right]\right.
\label{eq:msphere}\\
&&
\left.
\hspace{3cm}
+\sum_{J\geq 0}\,(-i)^J\sqrt{4\pi(2J+1)}\left[(\epsilon_a^*)_0\mathcal{L}_{J0}(k)
-\epsilon^{0\,*}_a\mathcal{M}_{J0}\right]
\right)
\vert i\rangle \ , \nnmb
\eeq
where
\beq
\mathcal{M}_{JM}(k) &=&
\int\!d^3x\,j_J(kr)Y_{JM}(\Omega){\mathcal{J}^0}(\vec{x})
\label{eq:scal}\\
\mathcal{L}_{JM}(k) &=& \frac{i}{k}\int\!d^3x\,\grad[j_J(kr)Y_{JM}(\Omega)]
\cdot\vec{\mathcal{J}}(\vec{x}) 
\label{eq:long}\\
\mathcal{T}^{el}_{JM}(k) &=&
\frac{1}{k}\int\!d^3x\;
\grad\times[j_{J}(kr)\yhat{J}{M}{J}(\Omega)]\cdot\vec{\mathcal{J}}(\vec{x})
\label{eq:tel}\\
\mathcal{T}^{mag}_{JM}(k) &=& \int\!d^3x\;
[j_{J}(kr)\yhat{J}{M}{J}(\Omega)]\cdot\vec{\mathcal{J}}(\vec{x}) \ .
\label{eq:tmag}
\eeq
The quantities $\yhat{J}{M}{\ell}$ are the \emph{vector spherical
harmonics}, defined according to~\cite{Blatt,Walecka:1995mi}
\beq
\yhat{J}{M}{\ell}(\Omega) = \sum_{m,\lambda}
\langle\ell m;1\lambda\vert\ell 1;JM\rangle\,
Y_{\ell m}(\Omega)\,\ehat_{\lambda} \ .
\eeq

The utility of the form of Eq.~\eqref{eq:msphere} is that the operators
appearing in the expansion, Eqs.~(\ref{eq:scal}--\ref{eq:tmag}), 
can be shown to be irreducible spherical tensors of degree $JM$
(for current operators of a reasonable form)~\cite{Walecka:1995mi}.  
This allows for the application of selection rules based on angular
momentum and parity.  In particular, for any such operator
$\mathcal{O}_{JM}$, the Wigner-Eckart theorem gives
\beq
\langle J_f,M_f\vert\mathcal{O}_{JM}\vert J_i,M_i\rangle
= \frac{(-1)^{J_i-M_i}}{\sqrt{2J+1}}\,
\langle J_f,M_f;\,J_i,-M_i\vert J_fJ_i;\,J,M\rangle\,
\langle J_f\Vert\mathcal{O}_{JM}\Vert J_i\rangle \ ,
\eeq
where the first matrix element refers to Clebsch-Gordan coefficients
and the second is a \emph{reduced matrix element} that
is independent of $M$, $M_i$, and $M_f$.  

For initial and final nuclear states of the form
$\vert i\rangle = \vert J_i,M_i\rangle$ and 
$\vert f\rangle = \vert J_f,M_f\rangle$, squaring and summing the
matrix element and applying the orthogonality of Clebsch-Gordan coefficients
gives
\beq
\frac{1}{2J_i+1}\sum_{M_i,M_f,a}|\mathcal{M}|^2 &=&
\frac{4\pi}{2J_i+1}\left(
\sum_{J\geq 1}\vert\langle J_f\Vert(\lambda\mathcal{T}_J^{mag}
+\mathcal{T}^{el}_J\Vert J_i\rangle\vert^2
\right.\label{eq:matvec}\\
&&\hspace{2cm}\left.
+\sum_{J\geq 0}\left[
\lrf{E_k}{m_X}^2\vert\langle J_f\Vert\mathcal{L}_J\Vert J_i\rangle\vert^2
+\lrf{k}{m_X}^2\vert\langle J_f\Vert\mathcal{M}_J\Vert J_i\rangle\vert^2
\right.\right.\nnmb\\
&&
\hspace{3.5cm}\left.\left.
-2\,\frac{kE_k^{\phantom{L}}}{m_X^2}\,\text{Re}\,\langle J_f\Vert\mathcal{L}_J\Vert J_i\rangle\langle J_f\Vert\mathcal{M}_J\Vert J_i\rangle^*
\right]
\phantom{\lrf{E}{E}^L}\hspace{-1.2cm}
\right)\nnmb \ .
\eeq
Note that this expression can also be adapted to decays to a massless
vector by setting $\mathcal{L}_J$ and $\mathcal{M}_J$ to zero, and to decays
to a scalar by keeping only $\mathcal{M}_J$ non-zero and removing the
factor of $(k/m_X)^2$ from the remaining term.

The final unpolarized decay rate for $\vert i\rangle \to \vert f\rangle + X$ 
(neglecting nuclear recoil effects) then follows from Fermi's Golden Rule~\cite{Walecka:1995mi}:
\beq \label{eq:Fermi}
\Gamma &=& \int\!\frac{d^3k}{(2\pi)^32E_k}\,(2\pi)\delta(M_i-M_f-E_k)\,
\frac{1}{2J_i+1}\sum_{M_i,M_f,a}|\mathcal{M}|^2\\
&=& \lrf{k}{2\pi}\,\frac{1}{2J_i+1}\sum_{M_i,M_f,a}|\mathcal{M}|^2 \ ,\nnmb
\eeq
where $k=\sqrt{(M_f-M_i)^2-m_X^2}$.
To evaluate this expression, the coupling current $\mathcal{J}^{\mu}(\vec{x})$
must be specified.

\subsection{Currents and Matrix Elements for an Axial Vector\label{sec:current}}

The hadronic current of Eq.~\eqref{eq:vcoup} to be used in the nuclear
matrix elements can be derived from quark- (and gluon-) level interactions
using the same methods as in dark matter direct detection 
studies~\cite{Engel:1992bf,Jungman:1995df,Fan:2010gt,Fitzpatrick:2012ix}.
In general, the fundamental quark-level interaction is matched onto 
an effective nucleon-level coupling based on chiral interactions.
Since the typical momenta relevant for nuclear decays are very small 
compared to the pion or nucleon masses, $k/m_N \sim 10^{-2}(k/10\,\mev)$,
the non-relativistic expansions used for dark matter calculations apply
to an excellent approximation.
These momenta are also much smallar than the inverse nuclear
radius $R^{-1}$, with
$kR \simeq 0.12\,({k}/{10}\,\mev)({A}/{8})^{1/3}$.  
Working to leading order in $k/m_N$ and $kR$, 
the general expression of Eq.~\eqref{eq:matvec} can be simplified considerably.

For an axial vector, we assume a coupling to quarks of the form
\beq
-\lag ~\supset~ X_{\mu}\,\sum_qg_q\bar{q}\gamma^{\mu}\gamma^5q \ ,
\eeq
where the sum runs over quark flavors.  When this operator is
inserted between a pair of nucleon states, the leading term
in an expansion in $k/m_N$ is~\cite{Engel:1992bf,Jungman:1995df,Agrawal:2010fh,Menendez:2012tm}
\beq
\langle N\vert \sum_qg_q\bar{q}\gamma^{\mu}\gamma^5q\vert N\rangle
= \delta^{\mu}_i\sigma^i\,\sum_qg_q\Delta q^{(N)} \ .
\eeq
The coefficients $\Delta q^{(N)}$ have been extrapolated from 
data~\cite{Mallot:1999qb,Ellis:2000ds,Cheng:2012qr} and computed
using lattice methods~\cite{QCDSF:2011aa,Engelhardt:2012gd,Abdel-Rehim:2013wlz,Chambers:2015bka,Green:2017keo}.
We use the recent combination of results in Ref.~\cite{Bishara:2016hek}:
\beq \label{eq:coeff}
\Delta u^{(p)} &=& \Delta d^{(n)} ~=~ ~~~0.897(27) \nnmb\\
\Delta d^{(p)} &=& \Delta u^{(n)} ~=~ -0.367(27) \\
\Delta s^{(p)} &=& \Delta s^{(n)} ~=~ -0.026(4)  \ , \nnmb
\eeq
where the proton-neutron equalities are expected to hold
to within the listed uncertainties.
The leading nucleon operator is often written in the 
isospin-inspired notation~\cite{Engel:1992bf}
\beq
-\lag_{eff} \supset \overline{N}(\vec{\sigma}\cdot\vec{X})
\frac{1}{2}(a_0+a_1\tau_3)N \ ,
\eeq
where $\tau_3$ is the Pauli matrix in isospin space and 
\beq 
a_0 &=& (\Delta u^{(p)}+\Delta d^{(p)})(g_u+g_d) + 2 g_s \Delta s^{(p)} \label{eq:a0}\\
a_1 &=& (\Delta u^{(p)}-\Delta d^{(p)})(g_u-g_d) \label{eq:a1} \ . 
\eeq
The corresponding forms for the proton and neutron are
$a_p = (a_0+a_1)/2$ and $a_n=(a_0-a_1)/2$.
From this, we can identify the leading-order current operator
to be used in nuclear matrix elements as~\cite{Walecka:1995mi}
\beq
\vec{\mathcal{J}}(\vec{x}) = \sum_{j=1}^Aa_j\vec{\sigma}^j\delta(\vec{x}-\vec{x}_j)
 \ ,~~~~~\mathcal{J}^0(\vec{x}) \to 0 \ ,
\eeq
where the sum runs over all nucleons.  The corresponding expression
for a (non-axial) vector can be found in Ref.~\cite{Walecka:1995mi}.

Turning next to nuclear matrix elements, the current operator derived
here can be applied to derive the transition operator in a spherical vector 
basis according to Eqs.~(\ref{eq:scal}--\ref{eq:tmag}).  The longitudinal
polarization of the massive vector gives non-zero $\mathcal{L}_{J0}$ terms, 
while the transverse polarizations lead to $\mathcal{T}_{J,\mp\lambda}^{mag,el}$ 
contributions.  However, to leading
order in $(kR) \sim 0.1$ this full machinery can be be bypassed
and the transition operator for an axial vector to be used 
in Eq.~\eqref{eq:M} simplifies to
\beq
\mathcal{O} &=& 
\int\!d^3x\,\sum_{\lambda}
e^{-i\vec{k}\cdot\vec{x}}\epsilon_{\lambda}^*\,
(\ehat_{\lambda}^*\cdot\mathcal{\vec{J}})
\label{eq:axop}\\
&=& \sum_{\lambda}\sum_{j=1}^A\,a_j\epsilon_{\lambda}^*\;
(\ehat^*_{\lambda}\cdot\vec{\sigma}^j ) + \mathcal{O}(kR)
\nnmb\\
&=& \sum_{\lambda}\sum_{j=1}^A\,a_j\epsilon_{\lambda}^*\;
(-1)^{\lambda}\sigma^j_{1,-\lambda} + \mathcal{O}(kR) \ ,
\nnmb
\eeq
where in the last line we have expressed $\vec{\sigma}$ as a spherical
tensor operator.

\subsection{Application to the Atomki Anomaly in $\beg$}

As an application of the above formalism, we turn next
to $\beg$ transitions related to the anomaly seen at the MTA-Atomki 
facility~\cite{Krasznahorkay:2015iga}.   
The relevant $\beg$ states, together with their properties, 
are listed in Table~\ref{tab:be8}.  Recall that an excess
bump-like feature is seen in the isoscalar $\bes(1^+)\to \beg(0^+)+e^+e^-$ mode,
but not in the related isovector $\bet(1^+)\to \beg(0^+)$ transition.
To evaluate whether the anomaly can be explained by a light axial
vector with $\bes(1^+) \to \beg(0^+)+X$, the isoscalar and isovector 
decay rates to the axial vector are needed.

The initial and final nuclear states in the $\bes\to \beg+X$
and $\bet\to \beg+X$ transitions have total angular momenta
$J_i = 1$ and $J_f = 0$, so the transition operator must be
a spherical tensor with $J=1$.  This implies
\beq
\langle J_f,M_f\vert\,\sigma_{1,-\lambda}^j\vert J_i,M_i\rangle
~\propto~ \delta_{M_i,\lambda} \ .
\eeq
Using this feature, we can use Eq.~\eqref{eq:axop} with the polarization
expressions of Eq.~\eqref{eq:eps} in Eq.~\eqref{eq:Fermi} to write 
the total decay width as
\beq \label{eq:width}
\Gamma = \frac{k}{6\pi}\,\left[
2\lvert\langle 0,0\vert\sum_{j=1}^Aa_j\sigma^j_{1,-}\vert 1,1\rangle\rvert^2
+ \lrf{E_k}{m_X}^2\lvert\langle 0,0\vert\sum_{j=1}^Aa_j\sigma^j_{1,0}\vert 1,0\rangle
\rvert^2
\right] \ .
\eeq
The sums in this expression can be split into neutron and proton pieces:
\begin{equation}
\sum_{j=1}^A a_j \sigma_{1,\lambda}^j = a_n\sum_{j=1}^{A-Z} \sigma_{1,\lambda}^{j,n} + a_p\sum_{j=1}^{Z} \sigma_{1,\lambda}^{j,p} \equiv 
a_n \hat{\sigma}^n_{1,\lambda} + a_p \hat{\sigma}^p_{1,\lambda}
\end{equation}
where the hatted operators signify the spin operators acting on all nucleons 
of a given type in the nucleus.  Using the Wigner-Eckart theorem, 
the various matrix elements can be written in terms of Wigner 3$j$ symbols 
and reduced matrix elements. 
This yields
\begin{equation}
 \langle 00| \hat{\sigma}_{1,-1}^{p,n} | 1 1 \rangle =  -\langle 00| \hat{\sigma}_{1,0}^{p,n} | 1 0 \rangle  = \frac{1}{\sqrt{3}} \langle 0 || \sigma^{p,n} || 1 \rangle.
\end{equation}
Inserting these expressions into Eq.~\eqref{eq:width} above, we have
\begin{equation}\label{eq:width2}
\Gamma = \frac{k}{18\pi}\,\left(2+\frac{E_k^2}{m_X^2}\right) \left|a_n \langle 0 || \sigma^n || 1 \rangle + a_p \langle 0 || \sigma^p || 1 \rangle \right| ^2 .
\end{equation}
Thus, the required nuclear input to the decay width consists of two reduced
matrix elements (for each of the two relevant $\beg$ excited states).

  The corresponding matrix element for electromagnetic transitions
must also have $J=1$.  Taking parity into account, it corresponds
to operators of the form $\mathcal{T}^{mag}_{J=1,\pm\lambda}$ 
in Eq.~\eqref{eq:msphere}.  For obvious reasons, these transitions
are referred to as $M1$~\cite{Blatt,Walecka:1995mi}.

\section{\emph{Ab Initio} Calculation of $^8$Be Matrix Elements\label{sec:nuc}}
 
To evaluate the nuclear matrix elements, we perform \emph{ab initio} 
calculations using realistic nuclear forces.  In the present case, this means
that we solve the full quantum mechanical system of eight nucleons (for $\beg$)
interacting with each other through forces derived from chiral effective field
theory using the in-medium similarity renormalization group~(IM-SRG)~\cite{Tsukiyama:2011,Hergert:2016,Hergert:2016a}, 
a recently-developed many-body method.

\subsection{Chiral Interactions}

  The inter-nucleon interactions used in our calculation are derived from 
chiral effective field theory and include two- and three-nucleon components.
For the two-nucleon~(NN) interaction, we use the result of Entem and Machleidt,
Ref.~\cite{Entem:2003}, derived at next-to-next-to-next-to-leading 
order~(N$^3$LO) in the chiral expansion, with a non-local regulator with cutoff $\Lambda_{NN}=500\,\mev$.
Importantly, this interaction includes the Coulomb force as well 
as nuclear isospin symmetry-breaking terms~\cite{Entem:2003}.
For the three-nucleon~(3N) interaction, we use the local N$^2$LO interaction 
of Navr\'atil, Ref.~\cite{Navratil:2007}, with cutoff $\Lambda_{3N}=400$ MeV 
\footnote{While the regulators used in the NN and 3N sectors are not the same,
there is no consensus as to how to consistently regulate the NN and 3N forces.
Fortunately, the present results are not sensitive to these details.}
and the two low energy constants $c_D$ and $c_E$ fit to the triton half-life 
and $A=3$ binding energies~\cite{Gazit:2009}.

To facilitate the convergence of the many-body calculation, the NN and 3N 
interactions are softened by the similarity renormalization group~(SRG) to a momentum scale $\lambda_{\text{SRG}}~=~2.0$~fm$^{-1}$~\cite{Bogner:2007,Roth:2014}.
We designate this interaction SRG~2.0.  As a check, we also employ the same 
interaction softened to a momentum scale $\lambda_{\text{SRG}}=1.88$~fm$^{-1}$.
Since the SRG is a unitary transformation (up to induced four-body forces),
the end results should be approximately independent of our choice 
of $\lambda_{\text{SRG}}$.
The lower cutoff $\Lambda_{3N}$ mentioned above was used in Ref.~\cite{Roth:2012}, and in many subsequent works (see e.g. \cite{Hergert:2013,Binder:2014,Soma:2014,Jansen:2014,Bogner:2014},
because -- in the region around $^{16}$O -- it produced results with a much weaker dependence on $\lambda_{\text{SRG}}$, indicating smaller induced 4N effects.
We also compare with calculations using the same N$^3$LO NN force but with the non-local N$^2$LO 3N interaction of 
Ref.~\cite{Hebeler:2011}, which is not consistently 
SRG-evolved, but instead has the 3N contact terms fit to reproduce the $^{3}$H binding energy and the $^{4}$He radius.
The NN force is SRG softened to $\lambda_{SRG}=1.8$~fm$^{-1}$, 
while the 3N force uses a regulator $\Lambda_{3N}=2.0$~fm$^{-1}\approx 395\,\mev$.
This interaction -- which was previously used to study nuclear matter~\cite{Hebeler:2011,Drischler:2016},
$sd$-shell nuclei~\cite{Simonis:2016}, and selected calcium~\cite{Hagen:2016,GarciaRuiz:2016} and nickel~\cite{Hagen:2016Ni78} isotopes
-- is designated EM~1.8/2.0.

\subsection{Many-Body Calculation}

We perform the many-body calculation using the IM-SRG, which we summarize below.
A more detailed review may be found in Ref.~\cite{Hergert:2016}.
In this method, the Hamiltonian
\begin{equation}
H = T_{\text{rel}} + V_{\text{NN}} + V_{\text{3N}} \ ,
\end{equation}
consisting of the relative kinetic energy plus the NN and 3N inter-nucleon 
interactions, is evaluated in a harmonic oscillator basis with frequency 
$\hbar\omega$. Since the harmonic oscillator eigenstates form a complete basis, 
an arbitrary wave function may be represented using an infinite number of
basis states, independent of the choice of $\hbar\omega$.  In our calculation,
we apply a single-particle truncation 
$2n+\ell \leq e_{max}$, where $n$ is the radial quantum number and $\ell$ 
is the orbital angular momentum, so that our results would become
exact in the limit $e_{max}\to \infty$.

Before implementing the IM-SRG, we begin by performing a spherical 
Hartree-Fock calculation of the $^{8}$Be ground state explicitly including 
the 3N interaction.  The interaction is then normal-ordered with respect 
to the Hartree-Fock ground state,\footnote{As discussed in Ref~\cite{Stroberg:2016a}, because we use a spherical formalism to treat an open-shell system, the reference is not a wave function but instead a particle-number violating ensemble, or mixed-state, reference. However the states produced in the final calculation are proper wave functions with good particle number.}
and the residual 3N force is discarded.
Note that while we discard the residual 3N piece, we retain most 
of the original 3N force through its normal-ordered 0-, 1-, and 2-body parts.
This approximation has been shown to be sufficient to capture the effects 
of 3N forces in the $p$-shell, such as the $1^+$-$3^+$ spin ordering 
in $^{10}$B~\cite{Stroberg:2016a,Gebrerufael:2016}.

Next, the IM-SRG is used to perform a unitary transformation $U$ which decouples 
a small valence space from the larger Hilbert space, producing an effective 
valence space interaction which approximately reproduces a subset 
of the eigenstates of the full space~\cite{Tsukiyama:2012,Bogner:2014}.
In the case of $^{8}$Be, we decouple the $0p$ shell model space.
To achieve this, we write the transformed Hamiltonian as~\cite{Morris:2015}
\begin{equation}
\begin{aligned}
\tilde{H} &= U H U^{\dagger} \\
&= e^{\Omega} H e^{-\Omega} \\
&= H + [\Omega,H] + \frac{1}{2!} [\Omega,[\Omega,H]] + \frac{1}{3!}[\Omega,[\Omega,[\Omega,H]]] + \ldots
\end{aligned}
\label{eq:BCH_transform}
\end{equation}
where the operator $\Omega=-\Omega^{\dagger}$ is the generator of the 
transformation, and the square brackets indicate a commutator.
While the last line in Eq.~\eqref{eq:BCH_transform} contains an infinite number 
of terms, arbitrarily high precision may be obtained with a finite number of 
terms for well-behaved transformations $U$.  The task is then to obtain 
an operator $\Omega$ that produces a decoupled Hamiltonian.
We achieve this by parameterizing $\Omega$ in terms of a flow parameter $s$
and an operator $\eta(s)$ that determines the direction of the flow, 
and integrating a flow equation
\begin{equation}
\begin{matrix}
e^{\Omega(s+ds)} = e^{\eta(s) ds}e^{\Omega(s)} \\  \Downarrow  \\
\Omega(s+ds) = \Omega(s) + \eta(s)ds + \frac{1}{2}[\eta(s),\Omega(s)] + \ldots \ ,
\end{matrix}
\label{eq:BCH_product}
\end{equation}
making use of the Baker-Campbell-Hausdorff formula.
We choose~\cite{White:2002}
\begin{equation}
\eta(s) \equiv \frac{1}{2} \tan^{-1}\left( \frac{2\tilde{H}_{od}(s)}{\Delta(s)} \right) - h.c.
\label{eq:def_eta}
\end{equation}
where $\Delta(s)$ is an energy denominator given by the difference 
of the expectation values of $H(s)$ for the bra and ket states, 
and the so-called off-diagonal part of the Hamiltonian, $\tilde{H}_{od}$, 
is the part we wish to suppress. In the present case it is given by those 
terms which connect valence space configurations to configurations outside 
the valence space.  The arctangent in Eq.~\eqref{eq:def_eta} is motivated by 
the solution of a two-level system, and ensures that no over-rotation 
is performed, even in the case of small denominators.
In Eqs.~(\ref{eq:BCH_transform},\ref{eq:BCH_product}), 
we retain only up to normal-ordered two-body operators.
This approximation, denoted IM-SRG(2), is the main approximation of the method 
and typically produces absolute binding energies within approximately 
1\% of the full solution.
Evidently, as the Hamiltonian is decoupled, 
$\tilde{H}_{od}$ is suppressed, $\eta(s)\to 0$, 
and $\Omega(s)$ approaches a fixed point.

At this point, the valence space forms a sub-block that is fully decoupled 
from the full Hilbert space.  We diagonalize $\tilde{H}$ in the valence space, 
using the shell model code NuShellX~\cite{Brown:2014} to obtain the final 
wave functions.
All transition operators $\mathcal{O}$ are consistently transformed 
using Eq.~\eqref{eq:BCH_transform} replacing $H\to \mathcal{O}$, 
as presented in Ref.~\cite{Parzuchowski:2016}, 
and are then evaluated with the shell model wave functions.

\subsection{Results for $^{8}$Be}
\begin{figure}[!ttt]
\begin{center}
\includegraphics[width=0.8\textwidth]{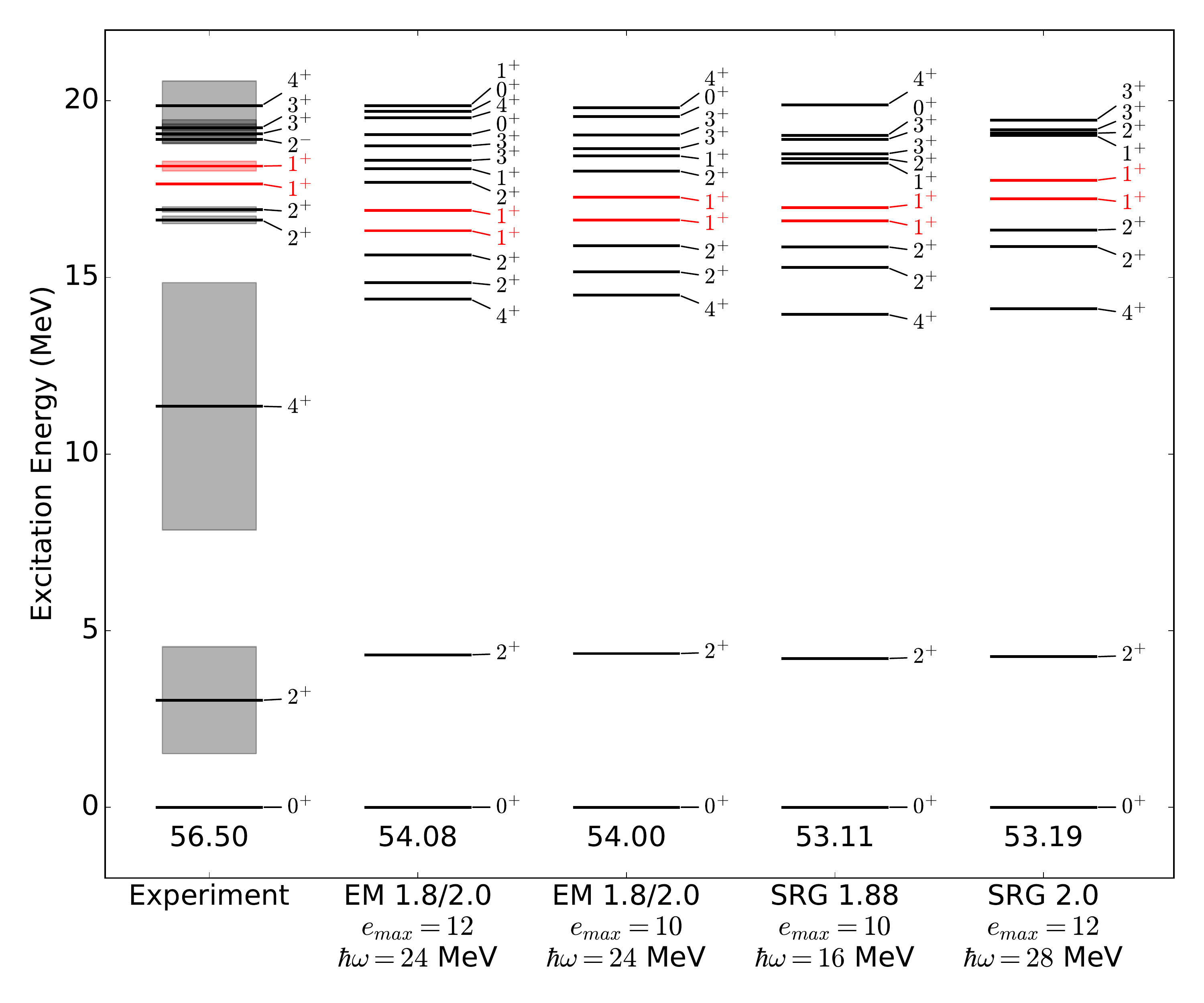}
\caption{\small \label{fig:spec} Experimental spectrum of $^{8}$Be labeled with total angular momentum and parity, compared with calculated spectra using the following interactions and model space parameters (see text for details). The shaded gray bands on the experimental spectrum indicate the width of the state. The $1^+$ states of interest are highlighted in red. Binding energies in MeV are also given beneath the ground states.} 
\end{center}
\end{figure}

In Fig.~\ref{fig:spec} we show the resulting excitation spectra in $\beg$
up to 20 MeV for a few selected combinations of interactions and model spaces,
as well as the experimentally measured spectrum.  We find a reasonable 
reproduction of the spectrum for all cases, noting that broad resonances 
such as the low-lying $2^+$ and $4^+$ states are typically poorly 
represented in a harmonic oscillator basis. Fortunately, the states of interest 
are the lowest two $1^+$ states, corresponding to $\bes$ and $\bet$ in 
Table~\ref{tab:be8}, which are narrow and reproduced well by the calculations.

  The specific quantities of interest for the present work are the transition
matrix elements relevant to Eq.~\eqref{eq:width2} involving the
$\bes$ and $\bet$ states.  An important factor in the description of these states,
and particularly for the transition rates, is their isospin content. 
For example, owing essentially to the opposite signs of the proton 
and neutron spin $g$ factors, isovector $M1$ (magnetic dipole) transitions 
dominate over isoscalar $M1$ transtions in $N=Z$ nuclei (see, e.g., Ref.~\cite{Eisenberg:1970}), 
while the opposite is true for an axial vector coupling.
This feature can be seen in the $M1$ photon transition rates $\Gamma_{\gamma}$
of the $\bes$ and $\bet$ states listed in Table~\ref{tab:be8},
which are  much
larger for the $\bet$ ($T\simeq 1$) state than the $\bes$ ($T\simeq 0$) state.
Note, however, that these isospin assignments are only approximate 
and each physical state is a mixture of isospin eigenstates.

  Isospin mixing in this context is delicately sensitive to the energy 
splitting between the two $1^+$ states, and to the isospin breaking terms 
in the interaction.  As a result, it is difficult to calculate this 
isospin mixing fraction with high precision.  However, since both the vector 
($M1$) and axial vector transition rates depend on the mixing, 
the two quantities become correlated.  We adopt the strategy used in 
Refs.~\cite{Hagen:2016,Calci:2016,Hagen:2016Ni78} to predict the axial vector matrix elements 
using their correlation with the isospin mixing and the known $M1$ transition 
strengths.\footnote{The isospin mixing fraction is not an observable quantity, but it is a useful heuristic to understand the correlation between the $M1$ and axial vector matrix elements. Using this correlation directly produces similar results.}

Let us denote the predominantly isoscalar $\bes$ and isovector $\bet$
states by $|\mathcal{S}\rangle$ and $|\mathcal{V}\rangle$, respectively,
and the pure isospin eigenstates by $|T=0\rangle$ and $|T=1\rangle$. 
Since our calculation methods violate isospin from the beginning, 
we do not have direct access to the pure isospin states.
Instead, we follow Ref.~\cite{Pastore:2014} and treat the isospin mixing 
as two-level mixing, so that the physical states are given by
\begin{equation}
\begin{aligned}
| \mathcal{S} \rangle &= ~~\beta    |T=0\rangle  + \alpha |T=1\rangle  \\
| \mathcal{V} \rangle &= -\alpha  |T=0\rangle  + \beta |T=1\rangle \ .
\end{aligned}
\end{equation}
The isospin mixing parameters may be obtained by 
\begin{equation}
\begin{matrix}
|\alpha|^2 = \frac{1}{2} \langle \mathcal{S} | \hat{T}^2 | \mathcal{S} \rangle
&,&
|\beta|^2 = \frac{1}{2} \langle \mathcal{V} | \hat{T}^2 | \mathcal{V} \rangle
&,&
\alpha\beta = \frac{1}{2} \langle \mathcal{S} | \hat{T}^2 | \mathcal{V} \rangle
\end{matrix}
\end{equation}
where $\hat{T}^2$ is the squared isospin operator.

Meson exchange currents~(MEC) in the nuclear current operators have 
not been included in our calculation.  The effect of MECs on $M1$ transitions 
in $^{8}$Be was investigated in Ref.~\cite{Pastore:2014} using a quantum 
Monte Carlo approach, yielding a 28\% correction to the isovector $M1$ 
matrix element.  To account for this, we correct the $M1$ matrix elements
obtained in our calculation by
\begin{eqnarray}
\delta_{MEC}(\mathcal{S}) &=& 0.28\,\left(\alpha^2 \langle \mathcal{S} \| M1 \| \mathcal{S} \rangle
 + \alpha\beta\langle \mathcal{V} \| M1 \| \mathcal{V} \rangle \right)
\\
\delta_{MEC}(\mathcal{V}) &=& 0.28\,\left(\beta^2 \langle \mathcal{V} \| M1 \| \mathcal{V} \rangle
 + \alpha\beta\langle \mathcal{S} \| M1 \| \mathcal{S} \rangle \right) \ .
\end{eqnarray}
The leading MEC correction to the axial current at low momentum 
is a two-body operator.
We follow Ref.~\cite{Menendez:2012} and treat the two-body 
contribution of this two-body operator by normal-ordering with respect 
to a Fermi gas.  This leads to a fractional correction to the isovector 
part of the current of
\begin{equation}
\delta a_1 = - \frac{\rho}{F_\pi^2} I(\rho,P=0)\left[ \frac{1}{3} (2c_4-c_3)+\frac{1}{6m_N} \right] \ ,
\end{equation}
where $\rho$ is the nucleon density, $F_\pi$ is the pion decay constant, 
$c_3$ and $c_4$ are low energy constants of the NN interaction, 
$m_N$ is the nucleon mass, 
and the quantity $I(\rho,P=0)$, defined as
\begin{equation}
I(\rho,P=0) = 1 - \frac{3m_\pi^2}{k_F^2} + \frac{m_\pi^3}{2k_F^3}\cot^{-1}\left(\frac{m_\pi^2-k_F^2}{2m_\pi k_F} \right) \ ,
\label{eq:Irho}
\end{equation}
is due to summation in the exchange term.
In Eq.~\eqref{eq:Irho}), $k_F=(3\pi^2\rho / 2)^{1/3}$ is the Fermi momentum 
of the Fermi gas, and $m_\pi$ is the pion mass.
Taking $\rho\approx 0.10$ fm$^{-3}$ yields $\delta a_1 \approx -0.25$.
We incorporate this fractional correction by scaling the proton axial 
vector matrix elements by $(1+\frac{1}{2}\delta a_1)$ and the neutron matrix 
elements by $(1-\frac{1}{2}\delta a_1)$.

\begin{figure}[!!t]
\begin{center}
\includegraphics[width=0.9\textwidth]{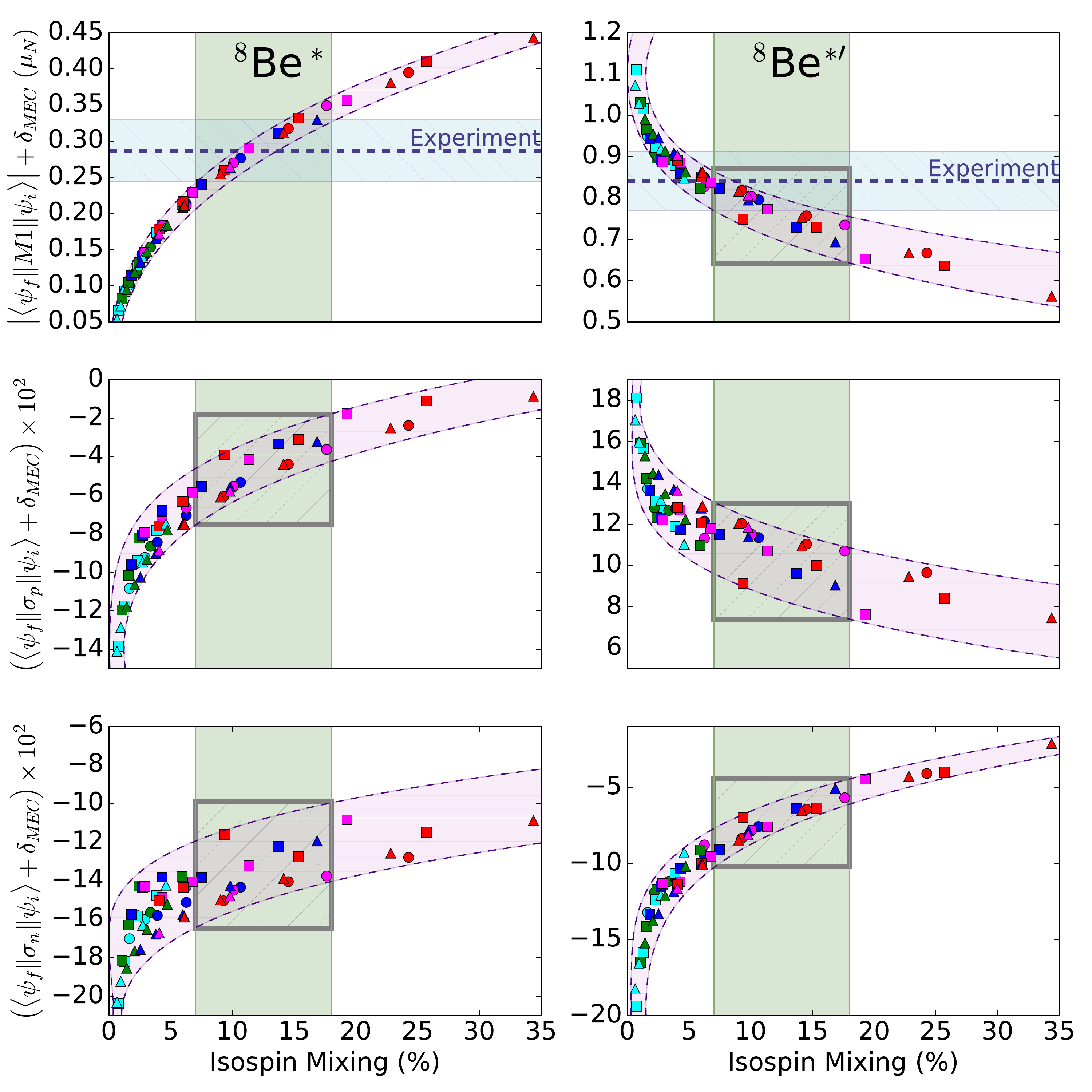}
\caption{\small  Reduced transition matrix elements for the $M1$, $\sigma_p$, 
and $\sigma_n$ operators between the $|\mathcal{S}\rangle$~($\bes$, left column)
and the $|\mathcal{V}\rangle$~($\bet$, right column) $1^+$  excited states and 
the ground state of $^{8}$Be as a function of the isospin mixing 
fraction $|\alpha|^2$.  Approximate corrections for meson exchange currents 
have been included. Circles indicate results using the SRG 1.88 interaction, 
triangles indicate the SRG 2.0 interaction, 
and squares indicate the EM 1.8/2.0 interaction. 
The single-particle basis truncations are indicated by different colors: 
$e_{max} = $  4~(cyan), 6~(green), 8~(blue), 10~(magenta), 12~(red).  
We include points for oscillator frequencies 
$\hbar\omega$=12, 16, 20, 24, and 28 MeV.
The $M1$ matrix element for the $T\simeq 0$, $J^P=1^+$ state in the upper 
left is used to constrain the range of the isospin mixing fraction,
which is then used to make predictions for the 
other matrix elements, indicated by the hashed boxes.}
\label{fig:sigpsign}  
\end{center}
\end{figure}

In Fig.~\ref{fig:sigpsign} we show the matrix elements of the $M1$ transition 
operator (corrected for MECs) and the proton and neutron spin operators 
$\sigma_p$ and $\sigma_n$ connecting the ground state to each of the 
lowest two $1^+$ states, calculated with the chiral interactions described above.
For each interaction, points are shown for a range of basis truncations 
$e_{max}$ and oscillator frequencies $\hbar\omega$, establishing a clear 
correlation between the matrix elements and the isospin mixing.
In the figures we multiply the $\sigma$ matrix elements by the sign of 
the $M1$ matrix element to eliminate effects due to the (arbitary) 
relative sign of the initial and final wave functions.

\begin{table}[ttt]
\centering
\begin{tabular}{c | D{.}{.}{3.9} }
\hline
\hline
Matrix element & \multicolumn{1}{c}{Prediction} \\
\hline
$\langle 0^+ \| M1       \| \mathcal{V} \rangle$ &  0.76 (12)~\mu_N \\
$\langle 0^+ \| \sigma_p \| \mathcal{V} \rangle$ &  0.102 (28) \\
$\langle 0^+ \| \sigma_n \| \mathcal{V} \rangle$ & -0.073 (29) \\
$\langle 0^+ \| \sigma_p \| \mathcal{S} \rangle$ & -0.047 (29) \\
$\langle 0^+ \| \sigma_n \| \mathcal{S} \rangle$ & -0.132 (33) \\
\hline
\hline
\end{tabular}
\caption{\label{tab:mat_el} Predicted nuclear matrix elements for the various transitions of interest, obtained by the correlation method described in the text. The predicted value of the $M1$ matrix element  for the
physical isovector-like state ($\mathcal{V}$) is consistent with the experimental value $0.84(7)\,\mu_N$.}
\end{table}

As the $|\mathcal{S}\rangle$ state is predominantly $T=0$, 
MEC corrections to the decay of this state are smaller and we expect 
this calculation to be more accurate than for the decay of the 
$|\mathcal{V}\rangle$ state.  In the upper left panel 
of Fig.~\ref{fig:sigpsign} we observe a strong correlation between the 
$\langle 0^+ | M1 | \mathcal{S}\rangle$ matrix element and the isospin mixing, 
indicated by the purple band.
We use this correlation and the experimentally known $M1$ strength 
to constrain the isospin mixing in our calculations, 
and find $|\alpha|=0.35(8)$.  This is larger than the value $\alpha=0.21(3)$ 
extracted in Ref.~\cite{Barker:1966} from a fit to data based on shell 
model calculations and a bare $M1$ operator, but consistent with 
$\alpha=0.31(4)$ obtained in Ref.~\cite{Pastore:2014} that does
include MEC corrections.
With this constraint, we make predictions for the other matrix elements, 
indicated by the hashed boxes in Fig.~\ref{fig:sigpsign}.
Our results are summarized in Table~\ref{tab:mat_el}.

\section{The $^8$Be Anomaly from an Axial Vector\label{sec:be8}}

Equipped with the nuclear transition matrix elements and the formalism 
described above, we can now address the Atomki $\beg$ 
anomaly~\cite{Krasznahorkay:2015iga} in terms of a light axial vector. 
Recall that the anomaly is seen in isoscalar $\bes\to \beg$
transitions, but not in isovector $\bet\to \beg$.
We find that this feature can arise naturally for decays 
to a light axial vector.

\subsection{Isoscalar $\bes\to \beg+X$ Transitions}

The original experimental paper reporting the $\beg$ anomaly also
provided an interpretation in terms of a light vector 
boson~\cite{Krasznahorkay:2015iga}.  The best fit mass and decay 
rate explaining the observed deviation from the predicted internal 
pair creation signal assuming $\textrm{BR}(X\to e^+e^-)=1$ 
were reported to be
\beq \label{eq:fit_1}
 m_X \simeq 16.7 \, {\rm MeV}, \qquad \frac{\Gamma_{\bes \to \beg \,X}}{\Gamma_{\bes \to \beg \,\gamma}} \simeq 5.8\times 10^{-6} \ ,
\eeq
with $\Gamma_{\bes\to\beg \, \gamma} \simeq (1.9\pm 0.4)\,\text{eV}$~\cite{Tilley:2004zz}.
Best-fit points for fixed higher masses were subsequently presented 
in Ref.~\cite{Feng:2016ysn}, citing a private communication 
with the authors of Ref.~\cite{Krasznahorkay:2015iga}. These are:
  \begin{equation}
  \begin{aligned} \label{eq:fit_2}
& m_X \simeq 17.3 \, {\rm MeV}, \qquad \frac{\Gamma_{\bes \to \beg \,X}}{\Gamma_{\bes \to \beg \,\gamma}} \simeq 2.3\times 10^{-6} \\
 &m_X \simeq 17.6 \, {\rm MeV}, \qquad \frac{\Gamma_{\bes \to \beg \,X}}{\Gamma_{\bes \to \beg \,\gamma}} \simeq 5.0\times 10^{-7} .
 \end{aligned}
 \end{equation}
It is likely that the overall fit to the data is worse for these higher 
masses~\cite{Krasznahorkay:2015iga}.  The best-fit mass and width 
for an axial vector may also differ due to the potentially slightly 
different angular distribution of $e^+e^-$ pairs relative to a purely 
vector coupling.  However, in both cases more information about the 
experimental apparatus and analysis would be needed to investigate 
these features in detail.

\begin{figure}[ttt]
\begin{center}
\includegraphics[width=0.5\textwidth]{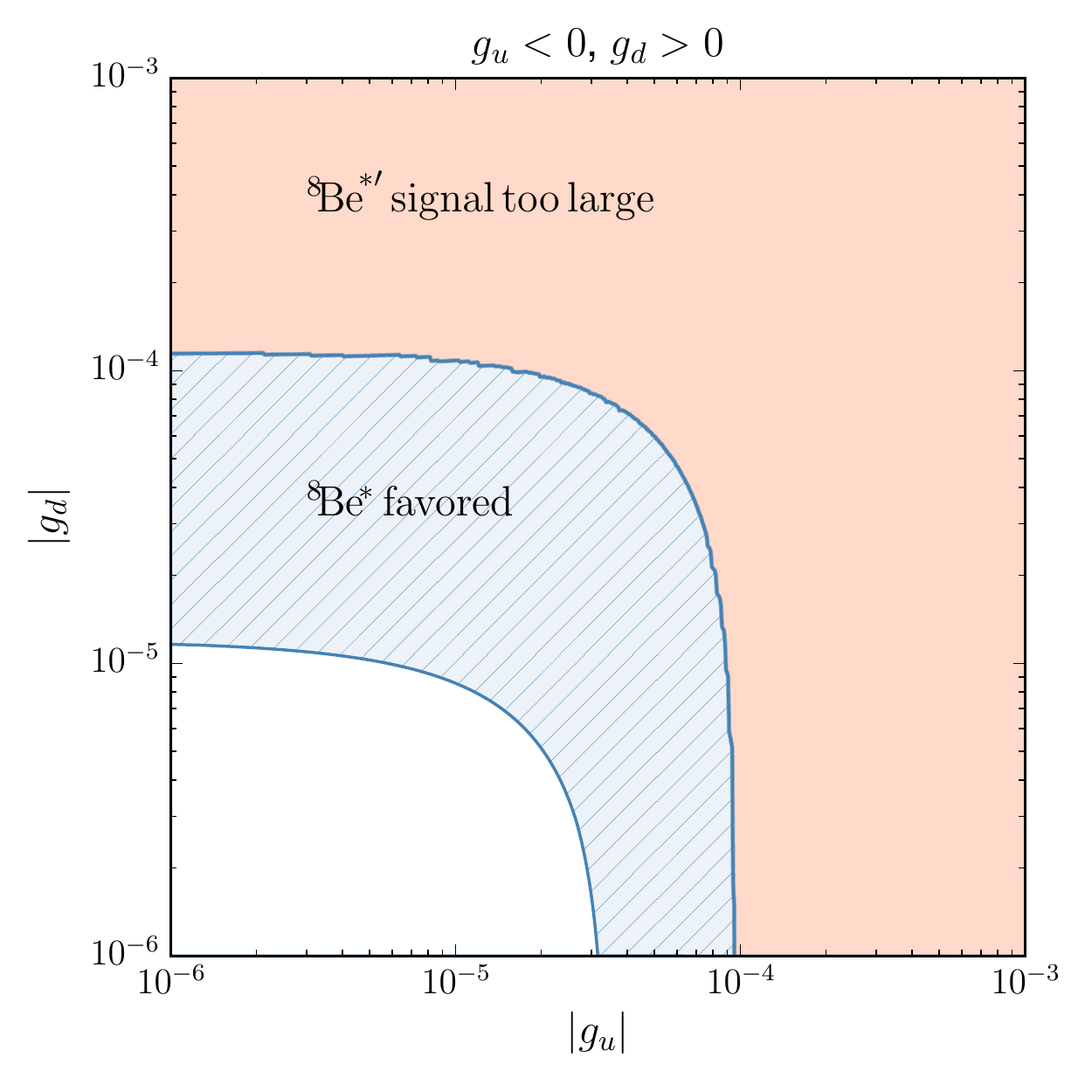}
\caption{\small \label{fig:couplings} 
Quark couplings required to explain the MTA-Atomki $^8$Be anomaly via 
a light axial vector assuming $g_u<0$, $g_d >0$, and electron couplings
such that $\mathrm{BR}(X\to e^+e^-)=1$ is prompt.  The hatched band was obtained by
considering  $m_X=16.7$, 17.3, 17.6 MeV, imposing the corresponding requirements in
Eqs.~(\ref{eq:fit_1})-(\ref{eq:fit_2}) to explain the MTA-Atomki result, then varying the relevant nuclear 
matrix elements and nucleon coefficients $\Delta u^{(p),(n)}$, $\Delta d^{(p),(n)}$ , $\Delta s^{(p),(n)}$ across their allowed ranges.  
Points below the hatched region feature
couplings too small to explain the observed $\bes$ transition rate for the three masses considered.
The orange region to the upper
right is excluded by the non-observation of an excess in the isovector 
$\bet\to \beg+e^+e^-$ channel for $m_X=16.7$-17.6 MeV.}
\end{center}
\end{figure}

Starting with the masses and decay widths listed above, we compute the range
of quark couplings to the axial vector that explain the $\beg$ anomaly.
To do so, we use Eqs.~(\ref{eq:a0},\ref{eq:a1}) to relate the quark
couplings $g_q$ to the coefficients $a_p$ and $a_n$, and then evaluate
the decay width of Eq.~\eqref{eq:width2} varying the nuclear matrix elements 
listed in Table~\ref{tab:mat_el}, as well as the nucleon coefficients in
Eq.~(\ref{eq:coeff}), across their uncertainty bands.  
The final results are shown in Fig.~\ref{fig:couplings} assuming
$g_u < 0$, $g_d >0$, $g_s=g_d$, and $\text{BR}(X \to e^+e^-)=1$.  

The ranges of potential axial vector quark couplings for the $\beg$ anomaly 
are fairly large due to the significant uncertainties on the values of 
the nuclear matrix elements. If the anomaly is confirmed in future experiments, 
it will be important to increase the precision of the nuclear calculation. 
Despite these uncertainties, we can draw some preliminary conclusions 
about the parameter space consistent with the anomaly.  In general, we find that  
$\operatorname{Max}(\left|g_u\right|, \left|g_d \right|) \gtrsim 10^{-5}$ 
is required to explain the result.  Note that this is significantly smaller
than the quark couplings needed for the protophobic vector explanation 
of the anomaly studied in Refs.~\cite{Feng:2016jff,Feng:2016ysn}.
This can be understood in terms of the leading partial wave for the decay,
with the axial vector decay proceeding at $\ell=0$ and proportional to 
$k/m_X \ll 1$ (from phase space), while the vector decay proceeds 
at $\ell = 1$ with a rate proportional 
to $k^3/m_X^3$~\cite{Feng:2016jff}.

\subsection{Isovector $\bet\to \beg+X$ Transitions}

The transition rate for $\bet\to \beg+X$ can be computed in the same
way as discussed above.  Since no significant excess was seen in
$\bet \to \beg+e^+ e^-$~\cite{Krasznahorkay:2015iga, Gulyas:2015mia}, 
we must check whether the quark couplings $g_q$ that explain the anomaly 
in the isoscalar channel are consistent with the data in the isovector mode.

  The condition we impose on the isovector channel for a given vector boson mass
follows that used in Ref.~\cite{Feng:2016jff}:
\beq \label{eq:isovector_bound}
 \frac{\Gamma_{\bes \to \beg \,X}}{\Gamma_{\bes \to \beg\,\gamma}}  > 5 \times \frac{\Gamma_{\bet \to \beg\,X}}{\Gamma_{\bet \to \beg\,\gamma}} \ .
\eeq
This (approximate) requirement is obtained by assuming that the statistical 
uncertainties on the $\bet$ transition are comparable to those 
for the $\bes$ transition, and that the ratios of the pair creation 
to electromagnetic transition rates are similar for both 
states.\footnote{We thank Jonathan Feng for clarification on this point.}. 
A more precise upper bound on the isovector transition rate would require
additional information about the MTA-Atomki detector sensitivities.

  In Fig.~\ref{fig:couplings} we show the impact of the $\bet$ condition 
of Eq.~\eqref{eq:isovector_bound}  on the possible ranges $g_u$ and $g_d$. 
Values of the couplings for which Eq.~\eqref{eq:isovector_bound} is not 
satisfied for any value of the nuclear matrix elements within the ranges quoted 
in Table~\ref{tab:mat_el}, nucleon coefficients within the ranges of Eq.~(\ref{eq:coeff}), 
and $m_X \in [16.7,\,17.6]\,\mev$ are indicated by
the orange shaded region in the figure.  The limit is the strongest 
model-independent constraint on the parameter space shown, highlighting 
the potential for nuclear decay experiments to probe previously unexplored 
theories of light vector bosons. The hatched region in Fig.~\ref{fig:couplings}
comprises the couplings that can be consistent with both the $\bes$ anomaly and the $\bet$ 
constraint.  Roughly, 
this requires $\operatorname{Max}(|g_u|, |g_d|)\lesssim 10^{-4}$.

The results of Fig.~\ref{fig:couplings} also reflect that
the $\bet\to \beg+X$ transition rate can be suppressed relative
to that of the $\bes \to \beg+X$ mode for an axial vector, which is
an important virtue of the axial vector interpretation.  
This effect is dynamical, as can be seen by comparing the relative sizes 
and signs of the reduced matrix elements in Table~\ref{tab:mat_el}.
In particular, the axial vector matrix elements are of similar size
for both the isoscalar and isovector states, while the $M1$ matrix
element relevant to the denominators in Eq.~\eqref{eq:isovector_bound}
is much larger for the isovector than the isoscalar.
This leads to a suppression of the isovector ratio 
in Eq.~\eqref{eq:isovector_bound} relative to the isoscalar 
that is not possible for a light gauge boson with only vector couplings, 
for which the relevant matrix elements are also proportional 
to those for the $M1$ transition.  One must then rely on kinematic 
suppression of the vector contribution to this transition by pushing 
the mass of the new particle closer to the $\bet$ 
threshold~\cite{ Feng:2016jff, Feng:2016ysn}, which appears to worsen 
the fit to experimental data.

\section{Constraints on Axial Vectors for the $^8$Be Anomaly \label{sec:constraints}}
 
In addition to the requirements on the quark couplings discussed above,
the axial vector must couple to leptons to allow it to decay 
to $e^+e^-$ pairs within the Atomki detector.  Lepton couplings also typically 
arise when the axial vector is embedded in a consistent UV-complete theory.  
Together, these quark and lepton couplings imply significant constraints on 
light vector explanations of the $\beg$ anomaly.  In this section 
we investigate the most significant constraints on a light vector
with axial quark couplings, making extensive use of the recent related 
analyses of Refs.~\cite{Fayet:2007ua,Feng:2016ysn, Kahn:2016vjr}.
These bounds will be applied to a UV complete theory of a light axial
vector in the section to follow.

To focus our study on the most important constraints on light axial vector 
explanations of the $^8$Be anomaly, we adopt the following assumptions:
\begin{enumerate}
\item The light vector $X$ has only axial couplings to quarks,
and these couplings are generation independent to avoid flavor mixing.  
\item Both vector and axial couplings to charged leptons are allowed
for the light vector: 
\beq
\mathcal{L}\supset X_\mu\,\sum_i \bar{\ell_i}\left(
g^V_{i}\gamma^{\mu} + g^A_{i} \gamma^{\mu} \gamma^5\right) \ell _i \ ,
\eeq
where the sums run over the charged leptons of the Standard Model.  
These couplings are again assumed to be generation independent.
\item The couplings of the vector boson to neutrinos vanish. 
This circumvents stringent constraints from electron-neutrino scattering 
experiments~\cite{Kahn:2016vjr,Bellini:2011rx, Deniz:2009mu, Vilain:1994qy}, 
and guarantees $\text{BR}(X\to e^+e^-)=1$ in the absence of 
other light states.
\end{enumerate}
With these assumptions, we compute the most significant constraints
on light vectors due their lepton and quark couplings.

\subsection{Lepton Coupling Constraints} 

For a light vector $X$ to explain the $^8$Be anomaly, 
its couplings to electrons must be large enough that it decays inside 
the Atomki detector. As pointed out in Refs.~\cite{Feng:2016ysn}, 
this implies
 \begin{equation}\label{eq:lepton_constraint}
 \frac{\sqrt{(g_e^V)^2 + (g_e^A)^2}}{e}  ~\gtrsim~ 1.3\times 10^{-5} \ .
 \end{equation}
Beyond this basic requirement, the lepton couplings of a light vector 
are constrained by lepton anomalous magnetic moments, beam dump searches, 
electron-positron collider experiments, 
and tests of parity violation in M\o ller scattering.

\subsubsection{Anomalous Magnetic Moments}

The anomalous magnetic moments of the charged leptons are affected by a 
light vector that couples to them. The corresponding shifts in
$a_{e,\mu} \equiv (g-2)_{e,\mu}$ for a general vector boson $X$ 
with both vector and axial couplings to leptons are~\cite{Fayet:2007ua}
\begin{equation}
\begin{aligned}
&\delta a_e = \frac{(g_e^{V})^2}{4\pi^2} \int_0^1\!\,dx\,\frac{x^2(1-x)}{x^2 + \frac{m_{x}^2}{m_e^2}(1-x)} - \frac{(g_e^{A})^2}{4\pi^2} \frac{m_e^2}{m_{x}^2} \int_0^1\!dx\,\frac{2x^3 + (x-x^2)(4-x)\frac{m_{x}^2}{m_e^2}}{x^2 + \frac{m_{x}^2}{m_e^2}(1-x)} 
\\
&\delta a_\mu = \frac{(g_\mu^V)^2}{4\pi^2} \int_0^1\!dx\,\frac{x^2(1-x)}{x^2 + \frac{m_{x}^2}{m_\mu^2}(1-x)} - \frac{(g_\mu^A)^2}{4\pi^2} \frac{m_\mu^2}{m_{x}^2} \int_0^1\!dx\, \frac{2x^3 + (x-x^2)(4-x)\frac{m_{x}^2}{m_\mu^2}}{x^2 + \frac{m_{x}^2}{m_\mu^2}(1-x)} \ .
\end{aligned}
\end{equation}
In general, the axial coupling of a light vector to leptons leads to negative 
contributions to their anomalous magnetic moments.  In the case of the muon, 
where the SM prediction is already lower than the measured value by about 
$3.4\,\sigma$~\cite{Bennett:2006fi,Blum:2013xva}, 
a light vector with purely axial couplings to muons worsens the disagreement.  

The interpretation of the measurement of 
$a_\mu$ as a constraint requires some care, since a naive application of 
the experimental result would also exclude the Standard Model. 
The disagreement between measurement and the SM prediction is 
about $2.9\pm 0.8 \times 10^{-9}$~\cite{Bennett:2006fi,Blum:2013xva}. 
To obtain a constraint from $a_\mu$, we demand that the contribution to
$\delta a_\mu$ from the axial vector be less than the $2\sigma$ uncertainty
(in either direction) of the discrepancy between experiment and the SM:
$\left|\delta a_\mu \right| \lesssim 1.6 \times 10^{-9}$. For $m_X\simeq 17$ MeV, 
this amounts to
\beq
\left|-(g_{\mu}^A)^2+9\times 10^{-3}(g_{\mu}^V)^2\right| \lesssim 1.6\times 10^{-9}.
\eeq
Let us also emphasize that numerous proposals have been made to explain
the disagreement in $a_\mu$, and many of them invoke weak-scale physics
that would not significantly alter the other low-energy observables 
considered here.  In this context, our requirement on $|\delta a_{\mu}|$ 
from a light axial vector corresponds to an absence of a strong cancellation 
with other contributions.

For the $a_e$ constraint, we impose
$-26\times 10^{-13} < \delta a_e \lesssim 8\times 10^{-13}$~\cite{Giudice:2012ms}.

\subsubsection{Electron Beam Dump Experiments}

Light vector bosons can be produced at electron beam dump 
experiments~\cite{Bjorken:2009mm}. For $m_X \simeq 17\,\mev$, 
the most stringent constraint comes from the SLAC E141 
experiment~\cite{Riordan:1987aw}, which 
requires~\cite{Alexander:2016aln,Kahn:2016vjr}
\begin{equation}
\frac{\sqrt{(g_e^{A})^{2} + (g_e^{V})^2}}{e}\gtrsim 2\times10^{-4}.
\end{equation}
In this regime, the vector $X$ would have decayed before reaching 
the detector. Other electron beam dump experiments yield less stringent bounds 
on the couplings; see Refs.~\cite{Feng:2016ysn, Kahn:2016vjr} for a more 
comprehensive discussion of these constraints.

\subsubsection{Electron-Positron Colliders}

A light axial vector coupling to electrons can be produced at $e^+ e^-$ 
experiments. The KLOE2 search for 
$e^+ e^- \to X \gamma$, $X\to e^+e^-$ 
sets a limit of~\cite{Feng:2016ysn, Kahn:2016vjr}
\begin{equation}
\frac{\sqrt{(g_e^{A})^{2} + (g_e^{V})^2}}{e}\lesssim2\times 10^{-3}
\end{equation}
for $m_X\simeq 17$ MeV. The BABAR experiment also searched for 
$e^+ e^- \to X \gamma$, 
$X\to \ell^+ \ell^-$, but only down to 
$m_X\gtrsim 20\,\mev$~\cite{Lees:2014xha}.

\subsubsection{Parity Violating M$\mathbf{\o}$ller Scattering}

Mixed axial-vector couplings of $X$ to leptons induces parity violation 
in M$\o$ller scattering. This was studied in the E158 experiment 
at SLAC~\cite{Anthony:2005pm}, and for $m_X \simeq 17\,\mev$ produces the 
constraint~\cite{Kahn:2016vjr}
\begin{equation}
\left|g_e^V g_e^A\right| \lesssim 1\times 10^{-8} \ .
\end{equation}  
Aside from $a_\mu$, this limit gives the the most stringent upper 
bound on lepton couplings in the UV-complete scenario we discuss below.

\subsection{Quark Coupling Constraints}

Light vector bosons can be constrained further if they couple to both
quarks and leptons, as required to explain the $\beg$ anomaly.
The two most important quark coupling constraints on this scenario, and given our assumptions,
 come from $\eta$ decays and proton beam dump experiments.

\subsubsection{Rare $\eta$ Decays}

New light particles can contribute to rare decays of the $\eta$ meson.  
As discussed in Refs.~\cite{Kahn:2016vjr, Masjuan:2015cjl}, the decay amplitude for
$\eta \to \mu^+\mu^-$ receives a new contribution from
the axial vector approximately proportional to $g_{\mu}^A(g_u+g_d- c g_s)$
that interferes with the SM contribution. 
Here $c$ is a real $\mathcal{O}(1)$ number that depends on the precise values 
of the $\eta-\eta^{\prime}$ mixing parameters used.  
This new contribution can produce a significant shift in the decay width 
for this mode relative to the SM alone, which agrees with data to within 
about $1\sigma$.  To determine the corresponding constraint, 
we evaluate the $\eta\to \mu^+\mu^-$ partial width following 
Ref.~\cite{Masjuan:2015cjl}, and demand that the 
net shift be less than the $2\sigma$ uncertainty on the SM prediction. 
This corresponds roughly to
\beq
\frac{g_{\mu}^A(g_u+g_d-1.5 g_s)}{(m_X/{\rm MeV})^2} \lesssim 4\times 10^{-10}.
\eeq
 Note that this differs slightly from the bound quoted in 
Ref.~\cite{Kahn:2016vjr} obtained using a different value for 
the $\eta-\eta^{\prime}$ mixing angle; the impact of this difference 
is negligible on the parameter space of interest.

\subsubsection{Proton Fixed Target Experiments}

Proton fixed target experiments also constrain the quark couplings 
of the vector, this time in combination with the electron couplings. 
In particular, the limits from the $\nu$-Cal I experiment at the IHEP U70 
accelerator provide bounds on $X$ production from bremsstrahlung off 
the proton beam~\cite{Blumlein:2013cua}. In dark photon models, 
the corresponding bound is $\epsilon^2 \lesssim 3.7\times 10^{-13}$ 
or $\epsilon^2 \gtrsim 2.5\times 10^{-9}$; very small couplings are
allowed because most of the dark photons decay before the detector,
while larger couplings imply that the dark photons decay well after.

To recast this constraint onto the axial vector scenario, 
we reinterpret the constraint of the $\nu$-Cal experiment on
the number of dark vector bosons $N_{\rm sig}$ produced by bremsstrahlung 
off the initial beam that decay inside the fiducial volume of the detector. 
This number is given by
\begin{equation}
N_{\rm sig} = N_{\rm tot}\,\eta\,\int\!dE_X\;\frac{dN}{dE_X}\,P(E_X)
\end{equation}
where $N_{\rm tot}$ is the total number of proton collision events, 
$P(E_X)$ is the probability for the vector to decay inside the detector, 
$dN/dE_X$ is the differential $X$ vector production rate per proton 
interaction, and $\eta$ is the efficiency of the detector. 
The probability $P(E_X)$ is given by
\begin{equation}
P(E_X) = \operatorname{exp}\left(-\frac{d_1 m}{c \tau |\vec{p}|}\right) - \operatorname{exp}\left(-\frac{d_2 m}{c \tau |\vec{p}|} \right),
\end{equation}
with $d_1 = 64\,\text{m}$ the distance from the beam dump to the front end 
of the detector, $d_2 = 87\,\text{m}$ the distance to the rear, 
and $\tau$ the $X$ lifetime. The expressions for $dN/dE_X$ found 
in Ref.~\cite{Blumlein:2013cua} can be carried over directly to the 
pure axial case with the replacement $e \epsilon \to a_p$.\footnote{
In the generalized Fermi-Williams-Weiz{\"a}cker method 
used in Ref.~\cite{Blumlein:2013cua} to derive the bounds, 
the Bremsstrahlung production cross-section for $X$ is proportional 
to the cross-section for the Compton-like process $p + \gamma^* \to p + X$ (see e.g.~Ref.~\cite{Bjorken:2009mm} for a more detailed discussion). 
Upon inspecting the squared matrix element $|\mathcal{M}|^2$ for 
the $2\to 2$ process in both the pure vector and axial vector case, 
and using the Dirac algebra and Dirac equation to commute the 
$\gamma^5$ factors, one finds that they are identical for both processes, 
with the replacement $e \epsilon \leftrightarrow a_p$.} 
Requiring $N_{\rm sig}$ to be smaller than the corresponding upper limit 
presented in Ref.~\cite{Blumlein:2013cua} constrains the couplings 
$a_p$, $g_e^V$, and $g_e^A$. These bounds are shown in Fig.~\ref{fig:UV} 
and are generally found to be less stringent than other constraints 
on the relevant parameter space.

\subsubsection{Comments on Other Constraints}

The NA48/2 experiment~\cite{Batley:2015lha} constrains the decay 
$\pi^0 \to \gamma X$, $X\to e^+ e^-$. 
The amplitude for this process is proportional to the axial anomaly 
trace factor and vanishes for purely axial quark-$X$ couplings, 
up to chiral-symmetry breaking effects proportional to light quark 
masses~\cite{Feng:2016ysn,Kahn:2016vjr,Georgi:1985kw}. 
Note that this constraint required vector explanations of the $^8$Be 
anomaly to be ``protophobic''.  This feature also implies that there are
no strong constraints from pion decay constraints in proton beam dump 
experiments~\cite{Blumlein:2011mv}.

A related potential constraint comes from the KLOE-2 search for 
$\phi \to \eta X$, $X\to e^+e^-$, which depends
specifically on the coupling of the light vector to the 
strange quark~\cite{Babusci:2012cr}.
However, since the $\phi$ has $J^P=1^-$ and the quark couplings of
the light vector conserve parity, the argument for $\pi \to \gamma X$
can be applied here to the extent that the internal structure of the $\phi$ 
can be neglected.  Even omitting this suppression, setting the axial form
factor to be of the same order as the vector form factor we find that the bound
imposed by this decay mode is subleading relative to the others considered
in this section.

 Other constraints on light vectors arise from atomic parity 
violation experiments~\cite{Porsev:2009pr} and limits on new particles 
from neutron-nucleus scattering~\cite{Barbieri:1975xy}.  
Atomic parity violation does not give a bound in the present case since we
consider an axial vector that conserves parity in the quark sector,
but it would be relevant away from the purely axial (or vector) limit.
Bounds from neutron-nucleus scattering are expected to be less important 
than in the vector case due to the decoherence induced by the coupling of 
$X$ to nucleon spin rather than a conserved charge.  Note as well
that these constraints are already subdominant in the pure vector case.

\section{A UV Completion for the $\beg$ Anomaly}\label{sec:UV}

As we have seen, the constraints on a light vector boson can depend on both
its lepton and quark couplings.  In contrast, the $\beg$ anomaly only
specifies a range of quark couplings (provided the decay of the vector
to electrons is fast enough).  However, both the quark and lepton
couplings of a light vector boson will typically be related
to each other in an underlying UV complete theory.  In this section,
we construct a simple UV completion of a light vector with exclusively
axial couplings to quarks that satisfies the basic assumptions listed
at the beginning of Section~\ref{sec:constraints}.  We also show that
the theory can explain the Atomki $\beg$ anomaly while maintaining consistency
with existing experimental searches.

\subsection{A Simple UV-Complete Theory}

There has been recent interest in building UV-complete anomaly-free theories 
of light axially-coupled vector bosons~\cite{Kahn:2016vjr,Ismail:2016tod}. 
We will focus on the model presented in Sec.~5.2 of Ref.~\cite{Kahn:2016vjr},
and defer a more detailed model-building effort to a future investigation.

Consider a dark $U(1)_{\rm RH}$ gauge theory with coupling $g_D$ under which
the right-handed SM fermions (e.g.~$u^c$, $d^c$, $e^c$) are charged. 
Denote the corresponding charge of the RH SM fermion $f^c$ as $q_{f}$. 
The charges are assumed to be the same for all three generations, 
with $q_d=q_e$ and the SM Higgs taken to be neutral under $U(1)_{\rm RH}$.
We include two dark Higgs fields, $H_{u,d}^{\prime}$, both neutral under 
the SM gauge group and with $U(1)_{\rm RH}$ charges $-q_u$, $-q_d$, respectively.
The $U(1)_{\rm RH}$ symmetry is spontaneously broken by non-zero vacuum 
expectation values for the dark Higgses, $v_{u,d}^{\prime}$.
In addition to the explicit charges, we allow for non-zero kinetic mixing 
$\epsilon$ between $U(1)_{\rm RH}$ and $U(1)_Y$. 
This setup, detailed in Ref.~\cite{Kahn:2016vjr}, generically gives rise 
to mixed vector and axial couplings of the massive $U(1)_{\rm RH}$ vector boson
$X$ to the charged Standard Model fermions (but not neutrinos).

In the scenario described above, the usual Standard Model Yukawa terms 
are forbidden by gauge invariance.  Following Ref.~\cite{Kahn:2016vjr}, 
Yukawa interactions for the SM fermions can be generated by introducing 
a set of heavy new vector-like $SU(2)_L$ doublet fermions $\Psi_f$ 
(and their conjugates, $\Psi^c_f$) with $U(1)_{\rm RH}$ charges $-q_{f}$ 
and vector-like masses $M$ 
(assumed to be the same for all $\Psi$ for simplicity). 
The charges of $\Psi_f$ under the SM gauge group are assumed to be 
the same as those of the corresponding left-handed SM fermion doublet.
We can introduce the interactions 
\beq
\begin{aligned}
\mathcal{L}_{\rm Yukawa}^{\rm UV} = &- H_{u}^{\prime} \Psi_{u}^c y^{\prime}_u Q -  H \Psi_{u} y_{u} u^c - H_{d}^{\prime} \Psi_{d}^c y^{\prime}_d Q - H^{\dagger} \Psi_{d} y_{d} d^c \\
& -H_{d}^{\prime} \Psi_{e}^c y^{\prime}_e  L -  H^{\dagger} \Psi_{e} y_{e} e^c
 + {\rm h.c.}
 \end{aligned}
\eeq 
where $y^{\prime}_{u,d,e}$ are generation-independent 3$\times$3 matrices, 
$y_{u,d,e}$ are proportional to the corresponding Standard Model 
Yukawa matrices, and $Q,L$ and $H$ are the Standard Model quark, 
lepton, and Higgs doublets, respectively. Upon integrating out the vector-like 
fermions, these interactions yield effective SM-like Yukawa couplings of the form
\beq
\mathcal{L}_{\rm Yukawa}^{\rm IR} = y_{u,\rm eff} H Q u^c + y_{d,\rm eff} H^{\dagger} Q d^c + y_{e,\rm eff}H^{\dagger}L e^c + {\rm h.c.},
\eeq
 where $y_{f, \rm eff} \equiv y_f y_f^{\prime} v_{u,d}^{\prime}/M$. 
Note that $M$ must be larger than about a TeV to avoid constraints 
from LHC searches on new vector-like quarks and leptons\footnote{As discussed in Ref.~\cite{Kahn:2016vjr},  $M$ cannot be arbitrarily large: obtaining the sizeable top quark Yukawa coupling for fixed values of the up-type axial coupling and $m_X$ places an upper bound on $M$ (assuming perturbatively small couplings in the matrices $y, y^{\prime}$). However, we find that $M$ can easily be in the multi-TeV range for $m_X\approx 17$ MeV, axial quark couplings $\lesssim 10^{-4}$, and $\sim \mathcal{O}(1)$ couplings in $y,y^{\prime}$. We therefore expect the corresponding constraints to be readily satisfied in the parameter space relevant for explaining the $\beg$ anomaly.}. 
In this construction, we have assumed the framework of Minimal Flavor 
Violation~(MFV), whereby $\Psi_f$ transforms as a triplet under 
the corresponding $SU(3)_f$ flavor subgroup, and new
contributions to flavor-changing neutral currents are suppressed. 

Given the assumptions above, the couplings of the massive $U(1)_{\rm RH}$ boson, 
$X$, to quarks are purely axial 
provided the following relation is satisfied:
\begin{equation}
g_D q_{u} \simeq -2 g_D q_{d} \simeq \frac{4}{3} e \epsilon \ .
\end{equation}
Matching to our previous notation, this implies
\begin{equation}
g_u=-2 g_d, \qquad g_{e,\mu}^A=g_d, \qquad g_{e,\mu}^V = 2 g_d
\label{eq:uvcoup}
\end{equation}
where $g_d$ can be treated as a free parameter. As discussed 
in Ref.~\cite{Kahn:2016vjr}, demanding purely axial couplings to quarks 
requires a tuning of $\epsilon$. Since our goal is simply to demonstrate 
that viable UV complete axial vector scenarios explaining the $\beg$ anomaly 
exist, we will not comment further on this issue here.

As it stands, the would-be $U(1)_{\rm RH}$ gauge symmetry is anomalous.  
This can be corrected by introducing additional fermions charged under
$SU(3)_c$, $U(1)_Y$ and $U(1)_{\rm RH}$ to cancel the anomalies. 
These new fermions, dubbed \emph{anomalons} in Ref.~\cite{Kahn:2016vjr}, 
are vector-like under the SM gauge groups but chiral under $U(1)_{\rm RH}$. 
They are assumed to obtain masses from the expectation values of 
the two dark Higgs fields, which will also contribute to the mass of 
the $X$ vector boson.  Since the anomalons carry color charge,
their masses must be larger than about a TeV to be consistent with LHC searches.
Demanding $m_X \simeq 17\,\mev$ then implies that~\cite{Kahn:2016vjr}
\begin{equation}\label{eq:anomalon}
\sqrt{(g_d)^2+(g_u)^2} \lesssim \left( \frac{y_\psi}{4 \pi}\right) 
\times 10^{-4} \ ,
\end{equation}
where $y_\psi$ is the Yukawa coupling of the anomalon fermions to 
the dark Higgses, assumed to be the same for both up- and down-type species. 

In this setup, the dark Higgs bosons are SM singlets and are 
weakly constrained, coupling to the visible sector either through 
the $X$ vector boson, loops of the new vector-like fermions, 
or Higgs portal-type interactions.  We therefore expect that there 
is enough freedom in the Higgs sector to straightforwardly satisfy 
the corresponding (highly model-dependent) 
constraints on the new Higgs scalars.

Note that the present construction could be modified to allow for $q_e\neq q_d$ 
by introducing another dark Higgs field. One could also envision 
a UV completion with generation-dependent couplings, 
along the lines discussed in Sec.~5.3 of Ref.~\cite{Kahn:2016vjr}, 
at the cost of additional tunings.
Such modifications could potentially open up additional parameter space 
for explaining the $\beg$ anomaly in terms of a light axially-coupled vector, 
but we do not pursue these directions further here.

\subsection{Constraints on the Theory and the $\beg$ Anomaly}

\begin{figure}[ttt]
\begin{center}
\includegraphics[width=0.5\textwidth]{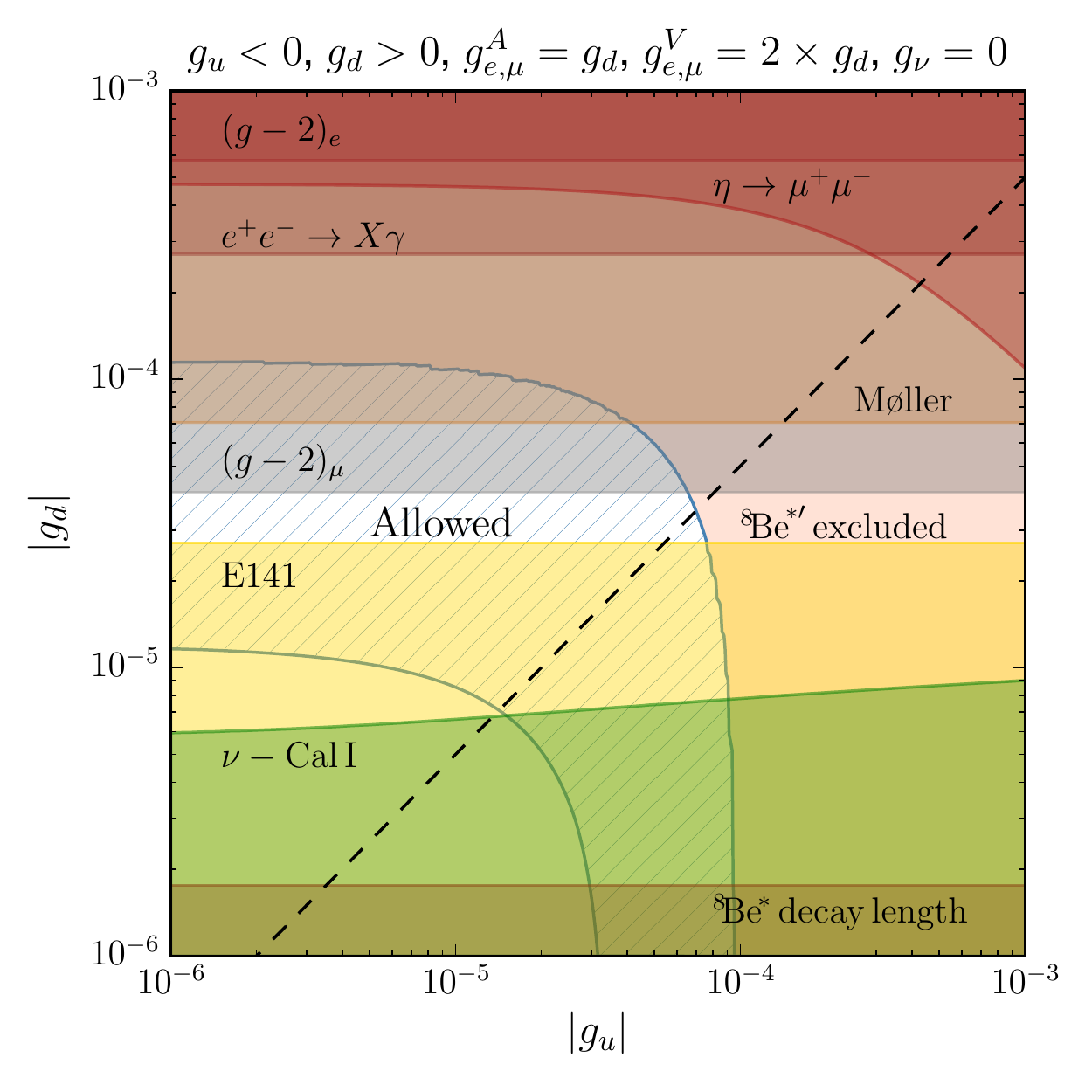}
\caption{\small \label{fig:UV} Quark level couplings required to explain the Atomki $^8$Be anomaly along with the most important constraints in the UV complete scenario described in Section~\ref{sec:UV}. For this specific model, $g_u$ and $g_d$ lie along the dashed black line. The experimentally allowed region is indicated as such, and includes values of the couplings consistent with an axial vector interpretation of the $\beg$ anomaly, depicted by the hatched region.} 
\end{center}
\end{figure}

Within this UV complete light axial vector scenario, we can now
connect the quark couplings needed to address the $\beg$ anomaly
to the many constraints on the theory that also depend on lepton couplings.
In Fig.~\ref{fig:UV} we show the most stringent bounds on the theory
in the $|g_u|$-$|g_d|$ plane with $g_u < 0$, $g_d > 0$, 
and the lepton couplings fixed in terms of $g_d$ as in Eq.~\eqref{eq:uvcoup}.  
Imposing the additional relation $g_u = -2g_d$ implied by the theory
gives the dashed diagonal line.  
We do not include the \emph{anomalon} 
bound of Eq.~\eqref{eq:anomalon} in the figure as the coupling $y_\psi < 4\pi$ can be chosen such it
does not constrain any additional parameter space. It would be beneficial to re-visit and sharpen
this bound in a more detailed phenomenological study, however we defer this to future work.
The hatched region in Fig.~\ref{fig:UV} indicates where the light vector 
can account for the Atomki $\beg$ anomaly. This band was obtained 
by varying $m_X$, the nuclear matrix elements in Table~\ref{tab:mat_el},
and the coefficients $\Delta u^{(p),(n)}$, $\Delta d^{(p),(n)}$ , $\Delta s^{(p),(n)}$ in Eq.~(\ref{eq:coeff}) 
across their allowed ranges, while also imposing the constraint
of Eq.~\eqref{eq:isovector_bound} on the $\bet$ transition rate. 
We see from this figure that there exists a small region of parameter
space (with $|g_d|\sim 3-4 \times 10^{-5}$) in which the light axial 
vector provides a viable explanation of the $\beg$ anomaly 
and is compatible with all other experimental constraints.

The strongest bounds on the theory tend to come from the lepton couplings
of the light vector.  Since these are fixed in terms of the quark couplings
by our choice of UV completion, it is possible that there are other consistent
UV models that are less constrained.  Even more parameter space could
open up if the assumptions about the couplings of the light vector 
listed at the start of Section~\ref{sec:constraints} were relaxed.
We postpone a more detailed investigation of these considerations to future work.

Let us also point out that the most important limit on the 
quark couplings alone comes from the Atomki measurements 
themselves~\cite{Krasznahorkay:2015iga}, with the entire region to 
the upper right of the hatched region in Fig.~\ref{fig:UV} excluded
by their data (up to nuclear uncertainties).  
Should the anomaly disappear in the future with more data, 
these constraints would become even stronger.  This again provides an
important illustration of how precision nuclear measurements 
can be used to study light vectors (and other particles) beyond
what is possible with other experiments.

\section{Conclusions\label{sec:conc}}

Rare nuclear decays are a promising search channel for new hidden particle 
species with masses near the MeV scale.  The anomaly seen in the 
$e^+e^-$ spectrum of isoscalar $\bes(1^+)\to \beg(0^+)$ transitions
at the Atomki facility can be explained by the emission of a light
vector boson in this process~\cite{Krasznahorkay:2015iga,Feng:2016jff}.  
In this paper, we have studied such an interpretation for a light vector 
boson with axial couplings to quarks.
To do so, we have performed a detailed \emph{ab initio} calculation
of the relevant nuclear transition matrix elements.  We find that such a
vector can account for the anomaly provided it has a mass of 
$m_X \simeq 17\,\mev$ and axial couplings to quarks 
on the order of $g_q \sim 10^{-5} - 10^{-4}$.
Relative to vector bosons with exclusively vector couplings to quarks,
the axial interpretation provides a natural suppression of vector
emission in the isovector $\bet(1^+)\to \beg(0^+)$ transition,
where no anomaly is seen.

In this work we have also investigated other constraints on light vector bosons
with axial quark couplings, and we have applied them to a simple UV realization
of the theory.  We find that the UV complete theory
studied here can explain the $\beg$ anomaly while being consistent with
all current experimental searches.  More generally, we also find that 
the Atomki measurements of the $\beg$ system can provide the most sensitive
model-independent probe of the interactions of a light vector with quarks.
This motivates future searches for light vector bosons and other particles
in rare nuclear transitions.

\section*{Acknowledgements}

We thank Sonia Bacca, Angelo Calci, Barry Davids, Jonathan Feng, Susan Gardner, Yonatan Kahn, 
Richard Hill, Jason Holt, Saori Pastore, Achim Schwenk, Johannes Simonis, Tim Tait, Flip Tanedo, and Richard Woloshyn for helpful comments and discussions. 
SRS would also like to thank Angelo Calci and Johannes Simonis for providing
the nuclear interactions used in this work.
The IM-SRG code used in this work employs the Armadillo 
library~\cite{Sanderson:2010}. Computations were performed with an allocation of computing resources at the JŸlich Supercomputing Center.
This work is supported by the Natural Sciences
and Engineering Research Council of Canada~(NSERC), with DM and SRS
supported in part by Discovery Grants. 
TRIUMF receives federal funding via a contribution agreement 
with the National Research Council of Canada.


\bibliographystyle{unsrt}

\begin{thebibliography}{9}

\bibitem{Leike:1998wr} 
  A.~Leike,
  Phys.\ Rept.\  {\bf 317}, 143 (1999)
  doi:10.1016/S0370-1573(98)00133-1
  [hep-ph/9805494].

\bibitem{Langacker:2008yv} 
  P.~Langacker,
  Rev.\ Mod.\ Phys.\  {\bf 81}, 1199 (2009)
  doi:10.1103/RevModPhys.81.1199
  [arXiv:0801.1345 [hep-ph]].

\bibitem{Carena:2004xs} 
  M.~Carena, A.~Daleo, B.~A.~Dobrescu and T.~M.~P.~Tait,
  Phys.\ Rev.\ D {\bf 70}, 093009 (2004)
  doi:10.1103/PhysRevD.70.093009
  [hep-ph/0408098].

\bibitem{Rizzo:2006nw} 
  T.~G.~Rizzo,
  hep-ph/0610104.

\bibitem{Aaboud:2016cth} 
  M.~Aaboud {\it et al.} [ATLAS Collaboration],
  Phys.\ Lett.\ B {\bf 761}, 372 (2016)
  doi:10.1016/j.physletb.2016.08.055
  [arXiv:1607.03669 [hep-ex]].

\bibitem{Khachatryan:2016zqb} 
  V.~Khachatryan {\it et al.} [CMS Collaboration],
  arXiv:1609.05391 [hep-ex].

\bibitem{Erler:2009jh} 
  J.~Erler, P.~Langacker, S.~Munir and E.~Rojas,
  JHEP {\bf 0908}, 017 (2009)
  doi:10.1088/1126-6708/2009/08/017
  [arXiv:0906.2435 [hep-ph]].

\bibitem{Borodatchenkova:2005ct} 
  N.~Borodatchenkova, D.~Choudhury and M.~Drees,
  Phys.\ Rev.\ Lett.\  {\bf 96}, 141802 (2006)
  doi:10.1103/PhysRevLett.96.141802
  [hep-ph/0510147].

\bibitem{Fayet:2007ua} 
  P.~Fayet,
  Phys.\ Rev.\ D {\bf 75}, 115017 (2007)
  doi:10.1103/PhysRevD.75.115017
  [hep-ph/0702176 [HEP-PH]].

\bibitem{Pospelov:2008zw} 
  M.~Pospelov,
  Phys.\ Rev.\ D {\bf 80}, 095002 (2009)
  doi:10.1103/PhysRevD.80.095002
  [arXiv:0811.1030 [hep-ph]].

\bibitem{Bjorken:2009mm} 
  J.~D.~Bjorken, R.~Essig, P.~Schuster and N.~Toro,
  Phys.\ Rev.\ D {\bf 80}, 075018 (2009)
  doi:10.1103/PhysRevD.80.075018
  [arXiv:0906.0580 [hep-ph]].

\bibitem{Batell:2009yf} 
  B.~Batell, M.~Pospelov and A.~Ritz,
  Phys.\ Rev.\ D {\bf 79}, 115008 (2009)
  doi:10.1103/PhysRevD.79.115008
  [arXiv:0903.0363 [hep-ph]].

\bibitem{Essig:2009nc} 
  R.~Essig, P.~Schuster and N.~Toro,
  Phys.\ Rev.\ D {\bf 80}, 015003 (2009)
  doi:10.1103/PhysRevD.80.015003
  [arXiv:0903.3941 [hep-ph]].

\bibitem{Reece:2009un} 
  M.~Reece and L.~T.~Wang,
  JHEP {\bf 0907}, 051 (2009)
  doi:10.1088/1126-6708/2009/07/051
  [arXiv:0904.1743 [hep-ph]].

\bibitem{Essig:2013lka} 
  R.~Essig {\it et al.},
  arXiv:1311.0029 [hep-ph].

\bibitem{Alexander:2016aln} 
  J.~Alexander {\it et al.},
  arXiv:1608.08632 [hep-ph].

\bibitem{Donnelly:1978ty} 
  T.~W.~Donnelly, S.~J.~Freedman, R.~S.~Lytel, R.~D.~Peccei and M.~Schwartz,
  Phys.\ Rev.\ D {\bf 18}, 1607 (1978).
  doi:10.1103/PhysRevD.18.1607

\bibitem{Treiman:1978ge} 
  S.~B.~Treiman and F.~Wilczek,
  Phys.\ Lett.\ B {\bf 74}, 381 (1978).
  doi:10.1016/0370-2693(78)90684-6

\bibitem{Savage:1986ty} 
  M.~J.~Savage, R.~D.~Mckeown, B.~W.~Filippone and L.~W.~Mitchell,
  Phys.\ Rev.\ Lett.\  {\bf 57}, 178 (1986).
  doi:10.1103/PhysRevLett.57.178

\bibitem{Hallin:1986gh} 
  A.~L.~Hallin, F.~P.~Calaprice, R.~W.~Dunford and A.~B.~Mcdonald,
  Phys.\ Rev.\ Lett.\  {\bf 57}, 2105 (1986).
  doi:10.1103/PhysRevLett.57.2105

\bibitem{Savage:1988rg} 
  M.~J.~Savage, B.~W.~Filippone and L.~W.~Mitchell,
  Phys.\ Rev.\ D {\bf 37}, 1134 (1988).
  doi:10.1103/PhysRevD.37.1134

\bibitem{Krasznahorkay:2015iga} 
  A.~J.~Krasznahorkay {\it et al.},
  Phys.\ Rev.\ Lett.\  {\bf 116}, no. 4, 042501 (2016)
  doi:10.1103/PhysRevLett.116.042501
  [arXiv:1504.01527 [nucl-ex]].

\bibitem{Tilley:2004zz} 
  D.~R.~Tilley, J.~H.~Kelley, J.~L.~Godwin, D.~J.~Millener, J.~E.~Purcell, C.~G.~Sheu and H.~R.~Weller,
  Nucl.\ Phys.\ A {\bf 745}, 155 (2004).
  doi:10.1016/j.nuclphysa.2004.09.059

\bibitem{Feng:2016jff} 
  J.~L.~Feng, B.~Fornal, I.~Galon, S.~Gardner, J.~Smolinsky, T.~M.~P.~Tait and P.~Tanedo,
  Phys.\ Rev.\ Lett.\  {\bf 117}, no. 7, 071803 (2016)
  doi:10.1103/PhysRevLett.117.071803
  [arXiv:1604.07411 [hep-ph]].

\bibitem{Feng:2016ysn} 
  J.~L.~Feng, B.~Fornal, I.~Galon, S.~Gardner, J.~Smolinsky, T.~M.~P.~Tait and P.~Tanedo,
  Phys.\ Rev.\ D {\bf 95}, no. 3, 035017 (2017)
  doi:10.1103/PhysRevD.95.035017
  [arXiv:1608.03591 [hep-ph]].

\bibitem{Gu:2016ege} 
  P.~H.~Gu and X.~G.~He,
  Nucl.\ Phys.\ B {\bf 919}, 209 (2017)
  doi:10.1016/j.nuclphysb.2017.03.023
  [arXiv:1606.05171 [hep-ph]].

\bibitem{Chen:2016dhm} 
  L.~B.~Chen, Y.~Liang and C.~F.~Qiao,
  arXiv:1607.03970 [hep-ph].

\bibitem{Liang:2016ffe} 
  Y.~Liang, L.~B.~Chen and C.~F.~Qiao,
  arXiv:1607.08309 [hep-ph].

\bibitem{Jia:2016uxs} 
  L.~B.~Jia and X.~Q.~Li,
  Eur.\ Phys.\ J.\ C {\bf 76}, no. 12, 706 (2016)
  doi:10.1140/epjc/s10052-016-4561-3
  [arXiv:1608.05443 [hep-ph]].

\bibitem{Kitahara:2016zyb} 
  T.~Kitahara and Y.~Yamamoto,
  Phys.\ Rev.\ D {\bf 95}, no. 1, 015008 (2017)
  doi:10.1103/PhysRevD.95.015008
  [arXiv:1609.01605 [hep-ph]].

\bibitem{Ellwanger:2016wfe} 
  U.~Ellwanger and S.~Moretti,
  JHEP {\bf 1611}, 039 (2016)
  doi:10.1007/JHEP11(2016)039
  [arXiv:1609.01669 [hep-ph]].

\bibitem{Chen:2016tdz} 
  C.~S.~Chen, G.~L.~Lin, Y.~H.~Lin and F.~Xu,
  arXiv:1609.07198 [hep-ph].

\bibitem{Kahn:2016vjr} 
  Y.~Kahn, G.~Krnjaic, S.~Mishra-Sharma and T.~M.~P.~Tait,
  JHEP {\bf 1705}, 002 (2017)
  doi:10.1007/JHEP05(2017)002
  [arXiv:1609.09072 [hep-ph]].

\bibitem{Seto:2016pks} 
  O.~Seto and T.~Shimomura,
  arXiv:1610.08112 [hep-ph].

\bibitem{Neves:2016ugb} 
  M.~J.~Neves and J.~A.~Helay\"el-Neto,
  arXiv:1611.07974 [hep-ph].

\bibitem{Blatt}
J.~M.~Blatt, V.~F.~Weisskopf, 
``Theoretical Nuclear Physics,''
Springer-Verlag, 1979 (864pp).

\bibitem{Walecka:1995mi} 
  J.~D.~Walecka,
  ``Theoretical nuclear and subnuclear physics,''
  Oxford Stud.\ Nucl.\ Phys.\  {\bf 16}, 1 (1995).

\bibitem{Engel:1992bf} 
  J.~Engel, S.~Pittel and P.~Vogel,
  Int.\ J.\ Mod.\ Phys.\ E {\bf 1}, 1 (1992).
  doi:10.1142/S0218301392000023

\bibitem{Jungman:1995df} 
  G.~Jungman, M.~Kamionkowski and K.~Griest,
  Phys.\ Rept.\  {\bf 267}, 195 (1996)
  doi:10.1016/0370-1573(95)00058-5
  [hep-ph/9506380].

\bibitem{Fan:2010gt} 
  J.~Fan, M.~Reece and L.~T.~Wang,
  JCAP {\bf 1011}, 042 (2010)
  doi:10.1088/1475-7516/2010/11/042
  [arXiv:1008.1591 [hep-ph]].

\bibitem{Fitzpatrick:2012ix} 
  A.~L.~Fitzpatrick, W.~Haxton, E.~Katz, N.~Lubbers and Y.~Xu,
  JCAP {\bf 1302}, 004 (2013)
  doi:10.1088/1475-7516/2013/02/004
  [arXiv:1203.3542 [hep-ph]].

\bibitem{Agrawal:2010fh} 
  P.~Agrawal, Z.~Chacko, C.~Kilic and R.~K.~Mishra,
  arXiv:1003.1912 [hep-ph].

\bibitem{Menendez:2012tm} 
  J.~Menendez, D.~Gazit and A.~Schwenk,
  Phys.\ Rev.\ D {\bf 86}, 103511 (2012)
  doi:10.1103/PhysRevD.86.103511
  [arXiv:1208.1094 [astro-ph.CO]].

\bibitem{Mallot:1999qb} 
  G.~K.~Mallot,
  Int.\ J.\ Mod.\ Phys.\ A {\bf 15S1}, 521 (2000)
  [eConf C {\bf 990809}, 521 (2000)]
  doi:10.1142/S0217751X00005309
  [hep-ex/9912040].

\bibitem{Ellis:2000ds} 
  J.~R.~Ellis, A.~Ferstl and K.~A.~Olive,
  Phys.\ Lett.\ B {\bf 481}, 304 (2000)
  doi:10.1016/S0370-2693(00)00459-7
  [hep-ph/0001005].

\bibitem{Cheng:2012qr} 
  H.~Y.~Cheng and C.~W.~Chiang,
  JHEP {\bf 1207}, 009 (2012)
  doi:10.1007/JHEP07(2012)009
  [arXiv:1202.1292 [hep-ph]].

\bibitem{QCDSF:2011aa} 
  G.~S.~Bali {\it et al.} [QCDSF Collaboration],
  Phys.\ Rev.\ Lett.\  {\bf 108}, 222001 (2012)
  doi:10.1103/PhysRevLett.108.222001
  [arXiv:1112.3354 [hep-lat]].

\bibitem{Engelhardt:2012gd} 
  M.~Engelhardt,
  Phys.\ Rev.\ D {\bf 86}, 114510 (2012)
  doi:10.1103/PhysRevD.86.114510
  [arXiv:1210.0025 [hep-lat]].

\bibitem{Abdel-Rehim:2013wlz} 
  A.~Abdel-Rehim, C.~Alexandrou, M.~Constantinou, V.~Drach, K.~Hadjiyiannakou, K.~Jansen, G.~Koutsou and A.~Vaquero,
  Phys.\ Rev.\ D {\bf 89}, no. 3, 034501 (2014)
  doi:10.1103/PhysRevD.89.034501
  [arXiv:1310.6339 [hep-lat]].

\bibitem{Chambers:2015bka} 
  A.~J.~Chambers {\it et al.},
  Phys.\ Rev.\ D {\bf 92}, no. 11, 114517 (2015)
  doi:10.1103/PhysRevD.92.114517
  [arXiv:1508.06856 [hep-lat]].

\bibitem{Green:2017keo} 
  J.~Green {\it et al.},
  arXiv:1703.06703 [hep-lat].


\bibitem{Bishara:2016hek} 
  F.~Bishara, J.~Brod, B.~Grinstein and J.~Zupan,
  JCAP {\bf 02}, 009 (2017)
  doi:10.1088/1475-7516/2017/02/009
  [arXiv:1611.00368 [hep-ph]].


\bibitem{Tsukiyama:2011}
 K.~Tsukiyama, S.~K.~Bogner, and A.~Schwenk,
 Phys.\ Rev.\ Lett.\ {\bf 106}, 222502 (2011)
 doi:10.1103/PhysRevLett.106.222502
 [arXiv:1006.3639 [nucl-th]]



\bibitem{Hergert:2016} 
  H.~Hergert, S.~K.~Bogner, T.~D.~Morris, A.~Schwenk and K.~Tsukiyama,
  Phys.\ Rept.\  {\bf 621}, 165 (2016)
  doi:10.1016/j.physrep.2015.12.007
  [arXiv:1512.06956 [nucl-th]].

\bibitem{Hergert:2016a} 
  H.~Hergert,
  arXiv:1607.06882 [nucl-th].


\bibitem{Entem:2003} 
  D.~R.~Entem and R.~Machleidt,
  Phys.\ Rev.\ C {\bf 68}, 041001 (2003)
  doi:10.1103/PhysRevC.68.041001
  [nucl-th/0304018].

\bibitem{Navratil:2007} 
  P.~Navratil,
  Few Body Syst.\  {\bf 41}, 117 (2007)
  doi:10.1007/s00601-007-0193-3
  [arXiv:0707.4680 [nucl-th]].

\bibitem{Gazit:2009}
  Doron~Gazit, Sofia~Quaglioni, and Petr~Navr\'atil,
  Phys.\ Rev.\ Lett. {\bf 103}, 102502 (2009)
  doi:10.1103/PhysRevLett.103.102502
  [nucl-th/0812.4444].

\bibitem{Bogner:2007} 
  S.~K.~Bogner, R.~J.~Furnstahl and R.~J.~Perry,
  Phys.\ Rev.\ C {\bf 75}, 061001 (2007)
  doi:10.1103/PhysRevC.75.061001
  [nucl-th/0611045].

\bibitem{Roth:2014} 
  R.~Roth, A.~Calci, J.~Langhammer and S.~Binder,
  Phys.\ Rev.\ C {\bf 90}, 024325 (2014)
  doi:10.1103/PhysRevC.90.024325
  [arXiv:1311.3563 [nucl-th]].

\bibitem{Roth:2012}
  Robert~Roth, Sven~Binder, Klaus~Vobig, Angelo~Calci, Joachim~Langhammer, and Petr~Navr\'atil
  Phys.\ Ref.\ Lett. {\bf 109}, 052501 (2012)
  doi:10.1103/PhysRevLett.109.052501
  [arXiv:1112.0287 [nucl-th]].


\bibitem{Hergert:2013}
  H.~Hergert, S.~Binder, A.~Calci, J.~Langhammer, and R.~Roth
  Phys.\ Rev.\ Lett.\ {\bf 110}, 242501 (2013)
  doi:10.1103/PhysRevLett.110.242501
  [arXiv:1302.7294 [nucl-th]]

\bibitem{Binder:2014}
  Sven Binder, Joachim Langhammer, Angelo Calci, and Robert Roth 
  Phys.\ Lett.\ B {\bf 736}, 119 (2014)
  doi:10.1016/j.physletb.2014.07.010
  [arXiv:1312.5685 [nucl-th]]

\bibitem{Soma:2014}
  V.~Som\`a, A.~Cipollone, C.~Barbieri, P.~Navrátil, and T.~Duguet
  Phys.\ Rev.\ C {\bf 89}, 061301(R) (2014)
  doi:10.1103/PhysRevC.89.061301
  [arXiv:1312.2068 [nucl-th]]


\bibitem{Jansen:2014}
  G.~R.~Jansen, J.~Engel, G.~Hagen, P.~Navr\'atil, and A.~Signoracci
  Phys.\ Rev.\ Lett.\ {\bf 113}, 142502 (2014)
  doi:10.1103/PhysRevLett.113.142502
  [arXiv:1402.2563 [nucl-th]]

\bibitem{Bogner:2014} 
  S.~K.~Bogner, H.~Hergert, J.~D.~Holt, A.~Schwenk, S.~Binder, A.~Calci, J.~Langhammer and R.~Roth,
  Phys.\ Rev.\ Lett.\  {\bf 113}, 142501 (2014)
  doi:10.1103/PhysRevLett.113.142501
  [arXiv:1402.1407 [nucl-th]].

\bibitem{Hebeler:2011} 
  K.~Hebeler, S.~K.~Bogner, R.~J.~Furnstahl, A.~Nogga and A.~Schwenk,
  Phys.\ Rev.\ C {\bf 83}, 031301 (2011)
  doi:10.1103/PhysRevC.83.031301
  [arXiv:1012.3381 [nucl-th]].

\bibitem{Drischler:2016}
  C.~Drischler, K.~Hebeler, and A.~Schwenk
  Phys.\ Rev.\ C {\bf 93}, 054314 (2016)
  doi:10.1103/PhysRevC.93.011302,
  [arXiv:1510.06728 [nucl-th]].


\bibitem{Simonis:2016}
  J.~Simonis, K.~Hebeler, J.~D.~Holt, J.~Men{\'{e}}ndez, and A.~Schwenk,
  Phys.\ Rev.\ C {\bf 93}, 011302 (2016)
  doi:10.1103/PhysRevC.93.011302,
  [arXiv:1508.05040 [nucl-th]].


\bibitem{Hagen:2016} 
  G.~Hagen {\it et al.},
  Nature Phys.\  {\bf 12}, no. 2, 186 (2015)
  doi:10.1038/nphys3529
  [arXiv:1509.07169 [nucl-th]].


\bibitem{GarciaRuiz:2016}
  R.~F.~Garcia Ruiz {\it et al.},
  Nature Phys.\ {\bf 12}, 594–598 (2016)
  doi:10.1038/nphys3645
  [arXiv:1602.07906 [nucl-ex]].

\bibitem{Hagen:2016Ni78}
  G.~Hagen, G.~R.~Jansen, and T.~Papenbrock,
  Phys.\ Rev.\ Lett.\ 117, 172501 (2016)
  doi:10.1103/PhysRevLett.117.172501
  [arXiv:1605.01477 [nucl-th]].


\bibitem{Stroberg:2016a} 
  S.~R.~Stroberg, A.~Calci, H.~Hergert, J.~D.~Holt, S.~K.~Bogner, R.~Roth and A.~Schwenk,
  Phys.\ Rev.\ Lett.\  {\bf 118}, no. 3, 032502 (2017)
  doi:10.1103/PhysRevLett.118.032502
  [arXiv:1607.03229 [nucl-th]].

\bibitem{Gebrerufael:2016} 
  E.~Gebrerufael, A.~Calci and R.~Roth,
  Phys.\ Rev.\ C {\bf 93}, no. 3, 031301 (2016)
  doi:10.1103/PhysRevC.93.031301
  [arXiv:1511.01857 [nucl-th]].


\bibitem{Tsukiyama:2012}
 K.~Tsukiyama, S.~K.~Bogner, and A.~Schwenk,
 Phys.\ Rev.\ C\ {\bf 85}, 061304(R) (2012)
 doi:10.1103/PhysRevC.85.061304
 [arXiv:1203.2515 [nucl-th]]


\bibitem{Morris:2015} 
  T.~D.~Morris, N.~Parzuchowski and S.~K.~Bogner,
  Phys.\ Rev.\ C {\bf 92}, no. 3, 034331 (2015)
  doi:10.1103/PhysRevC.92.034331
  [arXiv:1507.06725 [nucl-th]].

\bibitem{White:2002}
  Steven R. White,
  J. Chem. Phys. {\bf 117}, 7472 (2002)
  doi:10.1063/1.1508370

\bibitem{Brown:2014}
  B.~A.~Brown, and W.~D.~M.~Rae,
  Nucl.\ Data Sheets {\bf 120}, 115 (2014)
  doi:10.1016/j.nds.2014.07.022 

\bibitem{Parzuchowski:2016}
  N.~Parzuchowski, S.~R.~Stroberg, H.~Hergert, P.~Navr\'atil, and S.~K.~Bogner,
  in prep.; N.~Parzuchowski Ph.D. thesis, Michigan State University (2017).

\bibitem{Eisenberg:1970}
  J. Eisenberg and W. Greiner,
  ``Excitation Mechanism of the Nucleus''
  North Holland Publishing Co. (1970)


\bibitem{Calci:2016} 
  A.~Calci and R.~Roth,
  Phys.\ Rev.\ C {\bf 94}, no. 1, 014322 (2016)
  doi:10.1103/PhysRevC.94.014322
  [arXiv:1601.07209 [nucl-th]].

\bibitem{Pastore:2014} 
  S.~Pastore, R.~B.~Wiringa, S.~C.~Pieper and R.~Schiavilla,
  Phys.\ Rev.\ C {\bf 90}, no. 2, 024321 (2014)
  doi:10.1103/PhysRevC.90.024321
  [arXiv:1406.2343 [nucl-th]].

\bibitem{Menendez:2012}
  J. Men\'endez, D. Gazit, and A. Schwenk,
  Phys. Rev. D {\bf 86}, 103511 (2012)
  doi: 10.1103/PhysRevD.86.103511
  [http://link.aps.org/doi/10.1103/PhysRevD.86.103511].

\bibitem{Barker:1966}
  F.~C.~Barker,
  Nucl. Phys. A {\bf 83} 418 (1966).

  
  \bibitem{Gulyas:2015mia} 
  J.~Gulyas {\it et al.},
  Nucl.\ Instrum.\ Meth.\ A {\bf 808}, 21 (2016)
  doi:10.1016/j.nima.2015.11.009
  [arXiv:1504.00489 [nucl-ex]].

\bibitem{Bellini:2011rx} 
  G.~Bellini {\it et al.}, 
  Phys.\ Rev.\ Lett.\  {\bf 107}, 141302 (2011)
  doi:10.1103/PhysRevLett.107.141302
  [arXiv:1104.1816 [hep-ex]].
  
  \bibitem{Deniz:2009mu} 
  M.~Deniz {\it et al.} [TEXONO Collaboration],
  Phys.\ Rev.\ D {\bf 81}, 072001 (2010)
  doi:10.1103/PhysRevD.81.072001
  [arXiv:0911.1597 [hep-ex]].
  
\bibitem{Vilain:1994qy} 
  P.~Vilain {\it et al.} [CHARM-II Collaboration],
  Phys.\ Lett.\ B {\bf 335}, 246 (1994).
  doi:10.1016/0370-2693(94)91421-4
  
\bibitem{Bennett:2006fi} 
  G.~W.~Bennett {\it et al.} [Muon g-2 Collaboration],
  Phys.\ Rev.\ D {\bf 73}, 072003 (2006)
  doi:10.1103/PhysRevD.73.072003
  [hep-ex/0602035].

\bibitem{Blum:2013xva} 
  T.~Blum, A.~Denig, I.~Logashenko, E.~de Rafael, B.~Lee Roberts, T.~Teubner and G.~Venanzoni,
  arXiv:1311.2198 [hep-ph].

\bibitem{Giudice:2012ms} 
  G.~F.~Giudice, P.~Paradisi and M.~Passera,
  JHEP {\bf 1211}, 113 (2012)
  doi:10.1007/JHEP11(2012)113
  [arXiv:1208.6583 [hep-ph]].

\bibitem{Riordan:1987aw} 
  E.~M.~Riordan {\it et al.},
  Phys.\ Rev.\ Lett.\  {\bf 59}, 755 (1987).
  doi:10.1103/PhysRevLett.59.755
  
  \bibitem{Lees:2014xha} 
  J.~P.~Lees {\it et al.} [BaBar Collaboration],
  Phys.\ Rev.\ Lett.\  {\bf 113}, no. 20, 201801 (2014)
  doi:10.1103/PhysRevLett.113.201801
  [arXiv:1406.2980 [hep-ex]].
  
\bibitem{Anthony:2005pm} 
  P.~L.~Anthony {\it et al.} [SLAC E158 Collaboration],
  Phys.\ Rev.\ Lett.\  {\bf 95}, 081601 (2005)
  doi:10.1103/PhysRevLett.95.081601
  [hep-ex/0504049].

\bibitem{Masjuan:2015cjl} 
  P.~Masjuan and P.~Sanchez-Puertas,
  JHEP {\bf 1608}, 108 (2016)
  doi:10.1007/JHEP08(2016)108
  [arXiv:1512.09292 [hep-ph]].

  \bibitem{Blumlein:2013cua} 
  J.~Blumlein and J.~Brunner,
  Phys.\ Lett.\ B {\bf 731}, 320 (2014)
  doi:10.1016/j.physletb.2014.02.029
  [arXiv:1311.3870 [hep-ph]].
  
  \bibitem{Batley:2015lha} 
  J.~R.~Batley {\it et al.} [NA48/2 Collaboration],
  Phys.\ Lett.\ B {\bf 746}, 178 (2015)
  doi:10.1016/j.physletb.2015.04.068
  [arXiv:1504.00607 [hep-ex]].

\bibitem{Georgi:1985kw} 
  H.~Georgi,
  ``Weak Interactions and Modern Particle Theory,''
  Menlo Park, Usa: Benjamin/cummings ( 1984) 165p  

\bibitem{Blumlein:2011mv} 
  J.~Blumlein and J.~Brunner,
  Phys.\ Lett.\ B {\bf 701}, 155 (2011)
  doi:10.1016/j.physletb.2011.05.046
  [arXiv:1104.2747 [hep-ex]].

\bibitem{Babusci:2012cr} 
  D.~Babusci {\it et al.} [KLOE-2 Collaboration],
  Phys.\ Lett.\ B {\bf 720}, 111 (2013)
  doi:10.1016/j.physletb.2013.01.067
  [arXiv:1210.3927 [hep-ex]].

\bibitem{Porsev:2009pr} 
  S.~G.~Porsev, K.~Beloy and A.~Derevianko,
  Phys.\ Rev.\ Lett.\  {\bf 102}, 181601 (2009)
  doi:10.1103/PhysRevLett.102.181601
  [arXiv:0902.0335 [hep-ph]]. 

  \bibitem{Barbieri:1975xy} 
  R.~Barbieri and T.~E.~O.~Ericson,
  Phys.\ Lett.\  {\bf 57B}, 270 (1975).
  doi:10.1016/0370-2693(75)90073-8
  
\bibitem{Ismail:2016tod} 
  A.~Ismail, W.~Y.~Keung, K.~H.~Tsao and J.~Unwin,
  Nucl.\ Phys.\ B {\bf 918}, 220 (2017)
  doi:10.1016/j.nuclphysb.2017.03.001
  [arXiv:1609.02188 [hep-ph]].

  

\bibitem {Sanderson:2010}%
   {C.}~{Sanderson},
  Techical Report, NICTA {2010}%
[http://arma.sourceforge.net/armadillo{\_}nicta{\_}2010.pdf]  
  
 

\end{thebibliography}

\end{document}